\documentclass[aps,pre,twocolumn,groupedaddress]{revtex4-2}

\usepackage{amsmath}
\usepackage{graphicx}
\usepackage{bbm}
\usepackage{bbold}

\begin{document}

\title{Three-dimensional bipedal model with zero-energy-cost walking}

\author{Sergey \surname{Pankov}}
\affiliation{Harik Shazeer Labs, Palo Alto, CA 94301}
\date{\today}

\begin{abstract}

We study a three-dimensional articulated rigid-body biped model that possesses zero cost of transport walking gaits. Energy losses are avoided due to the complete elimination of the foot-ground collisions by the concerted oscillatory motion of the model's parts. The model consists of two parts connected via a universal joint. It does not rely on any geometry altering mechanisms, massless parts or springs. Despite the model's simplicity, its collisionless gaits feature walking with finite speed, foot clearance and ground friction. The collisionless spectrum can be studied analytically in the small movement limit, revealing infinitely many periodic modes. The modes differ in the number of sagittal and coronal plane oscillations at different stages of the walking cycle. We focus on the mode with the minimal number of such oscillations, presenting its complete analytical solution. We then numerically evolve it toward a general non-small movement solution. A general collisionless mode can be tuned by adjusting a single model parameter. Some of the presented results display a surprising degree of generality and universality. 

\end{abstract}


\maketitle

\newcommand{\T}{\mathsf T}

\section{Introduction}

Arguably, energy efficiency is one of the two most important issues in the field of robotic locomotion \cite{post2013robustness} (the other one -- control robustness -- is outside the scope of this paper). Efficiency makes an autonomous robot proportionally more useful, as it can run longer on a single power refill \cite{bhounsule2014low} (e.g. battery charge) and wears out its hardware more slowly. A common measures of efficiency for mobile robots is the cost of transport (COT) -- the amount of energy spent by a traveling robot per its weight per distance traveled.

The importance of energy efficiency is not limited to man-made machines. As it offers a clear survival advantage, it is natural to expect for animal gaits to be significantly shaped by the energy efficiency requirements \cite{alexander1989optimization}. Indeed, optimizing a detailed human neuromusculoskeletal model for COT produces a gait similar to a natural human gait \cite{anderson2001dynamic}.

Likewise, when anthropomorphic robots are designed with COT optimality as the main objective, their walking gaits often appear humanlike \cite{mcgeer1990passive,collins2001three}. Walking motion patterns in these robots are mostly decided by their passive dynamics. To sustain walking they need a relatively small energy injection, either through joint actuation or by walking on a slight incline. 

Since COT optimality plays a prominent role, it is reasonable to ask: how energy is lost during walking and how these losses can be minimized? In this work, we are only concerned with mechanical losses. In general, energy is lost when the robot's actuators work against its passive dynamics and when its feet interact with the ground. For a passive walker only the latter is of concern: energy is lost when a foot either collides with or slides against the ground. To avoid the collision loss the foot velocity must vanish at the contact. For the foot then to stay on the ground, its acceleration must also vanish \cite{reddy2001passive}, while the jerk (time derivative of acceleration) should remain finite at the contact \cite{chatterjee2002persistent}. This type of collisionless motion has been proposed and demonstrated in a hopper \cite{reddy2001passive,chatterjee2002persistent} and a rimless wheel \cite{gomes2005collisionless,gomes2015quiet}.

Note, if only velocity vanishes at the contact \cite{gomes2011walking}, an additional mechanism (e.g. suction cups) is needed to keep the robot in contact with the surface, as the surface reaction force is negative (directed into the surface away from the robot) when the contact occurs, as in brachiation \cite{gomes2005five}. If the ground reaction force is non-negative everywhere on a walking trajectory, we call such walking conventional.

For an appropriately designed walker, the collision losses can be made to scale as the fourth power of velocity \cite{garcia2000efficiency}, thereby COT vanishing as the third power at low velocity. From a certain practical stand point, such elimination of COT in the limit may be of limited utility. In this work we are concerned with finding a finite-speed collisionless solution. We are also not interested in nonphysical solutions, such as considering massless springs, even though they trivially provide for a lossless locomotion.

Some passive walker models require additional active mechanisms to practically realize their walking modes. For example, leg retraction may be needed to prevent ground scuffing \cite{garcia2000efficiency}, or knees may need to be actively locked for a portion of a walking cycle \cite{trifonov2007active}. While technically realizable, they add to the engineering challenge \cite{mcgeer1990passiveknees}, and ideally should be avoided, if possible.

In this work we report a three-dimensional passive walker model with a perfectly collisionless finite-speed walking gait. The model is free of nuisances requiring engineering intervention. The ground scuffing is avoided by rocking motion in the coronal plane, without altering the model's geometry. There is no joint locking. The robot is made of articulated rigid parts without use of springs. The walking gait is realized at finite ground friction. The existence of a collisionless solution is demonstrated analytically.
To date, to the best of our knowledge, no collissionless finite-speed conventional bipedal walking in a physically realistic model has been conclusively demonstrated by solving equations of motion, either numerically or analytically, in either two or three dimensions.

The analytical solution is obtained via a perturbation theory in the feet separation parameter, to the lowest order necessary to establish a nontrivial (i.e. walking) collisionless solution. This amounts to considering the sagittal and coronal dynamics to linear order and the axial dynamics to quadratic order. The perturbative treatment is justified under certain conditions, which we collectively term the small movement limit (SML). A SML solution can be numerically evolved to a general solution using constrained gradient descent optimization.

We showed that the number of model parameters needed for tuning a collisionless solution is in general independent of the model complexity and only depends on the topological properties of the foot-ground interaction. Interestingly, our SML solution is to a significant degree encoded by a pair of universal functions, completely independent of the model parameterization. It is an intriguing possibility that this universality of the collisionless solution is not coincidental to our model, but may be generalizable to other models as well.

The solution of our springless model features a peculiar hanging torso gait. To enable a more-anthropomorphic looking standing torso gait, we also considered a model endowed with springs. We established an up-down torso duality and exploited it to formulate the standing torso solution in terms of the original springless model solution.

The paper is organized as follows. The biped model is introduced in Sec. \ref{modelnotations}. In Sec. \ref{eqmots} we derive the exact equations of motion, as well as their approximate and simpler SML form. Certain symmetry constraints are imposed on the general solution in Sec. \ref{solutionconstrs}, to make the solution search more manageable. The SML solution is analytically investigated throughout Sec. \ref{smlsolution}. The found SML solution is employed as an initial guess for a general numerical solution presented in Sec. \ref{gensol}. In Sec. \ref{discussion} we discuss related work and outline directions for possible future research. Throughout this paper, for the sake of both clarity and completeness of presentation, many technical detailed have been relegated to the appendices.

\section{Model, terminology and notations}
\label{modelnotations}

In this section we introduce the biped model, related terminology and some ubiquitous notations.

\begin{figure}
  \includegraphics[width=.7\columnwidth]{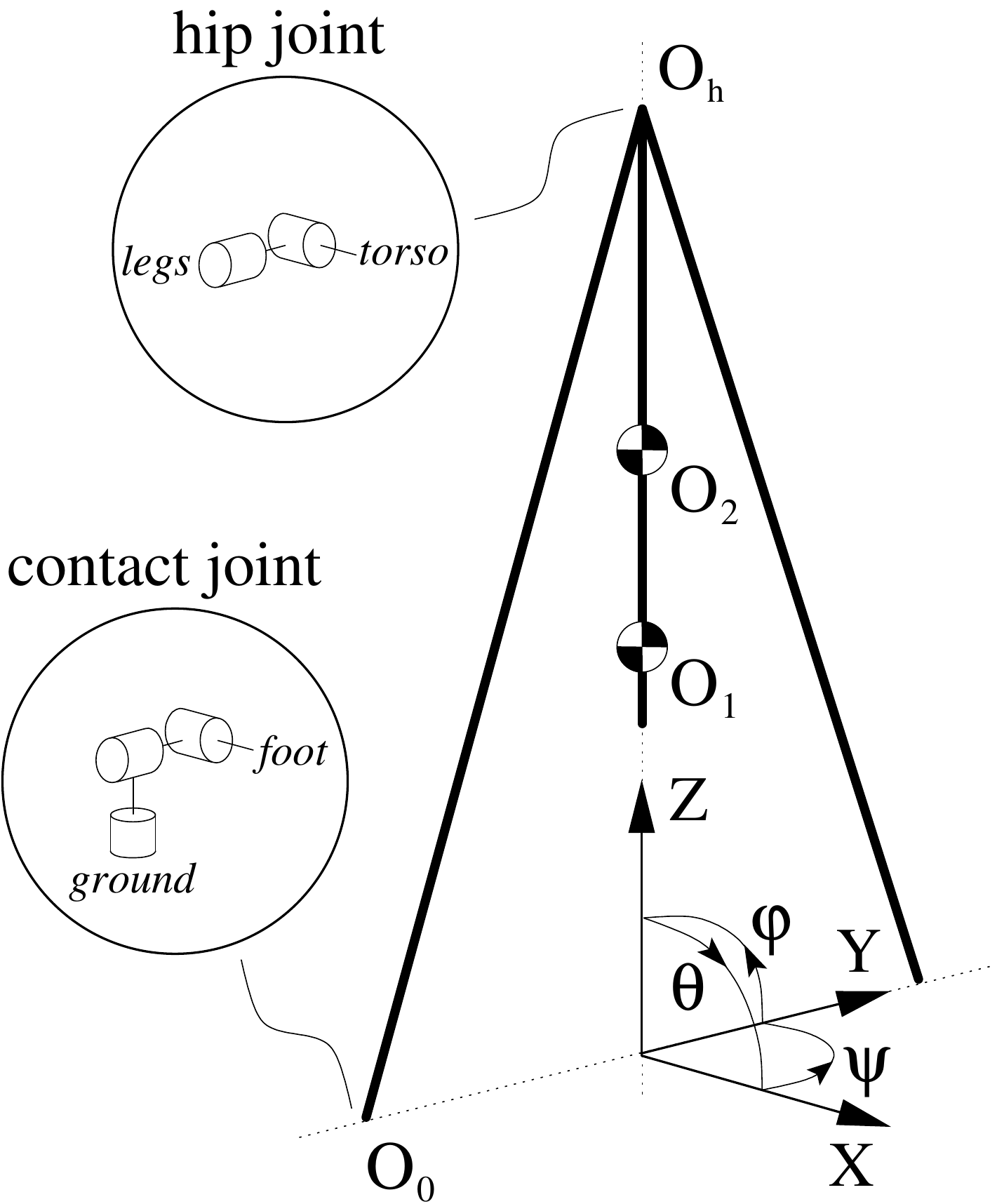}
  \caption{Biped model. The figure is schematic: in general, the depicted leg and torso links should not be regarded as indicative of the mass distribution.}
  \label{modelfig}
\end{figure}

The biped model is depicted in Fig.(\ref{modelfig}). It is composed of two rigid parts, called legs and torso, connected at the hip via a universal joint. The endpoints of the legs coming in contact with the ground ($z=0$ plane) are called feet. The hip and feet form an isosceles triangle (with the hip at the apex). In the figure, the biped is shown in the upright configuration, characterized by the reflection symmetries relative to the coronal ($x=0$) and sagittal ($y=0$) planes.

We limit our consideration to periodic walking motion. 
Walking consists of alternating single and double support phases. In the double support phase, called phase II, both feet are on the ground. In the single support phase, called phase I, only one foot (called stance foot) is on the ground. The other foot (called swing foot) is moved to a new location during phase I, thereby realizing a stepping motion.

Unless stated otherwise, by collisionless walking (gait, or solution) we mean collisionless conventional walking, that is with both velocity and acceleration vanishing at the moment of contact. Because collisionless gait conserves energy, it is time reversible. Therefore, we do not need to distinguish between a foot strike and lift-off. We refer to both as an impact event, foot impact or impact for short.

Before we proceed, a few words are in order about our matrix conventions. We use square brackets to denote a matrix, commas to separate matrix elements and semicolons to separate matrix rows. Vectors are in column format. Due to heavy use of sub- and superscripts, we use parentheses to explicitly indicate vector or matrix component indices, unless noted otherwise. A component index can be replaced by the dot symbol, to select all elements it represents. For example, $M_{(i\cdot)}$ and $M_{(\cdot j)}$ are $i$-th row and $j$-th column of a matrix $M$. Transposition is denoted $M^\T_{(ij)} = M_{(ji)}$. The Kronecker delta is $\delta_{ij} \equiv \delta_{(ij)}$. An identity matrix $I_{n\times m}$ is $n\times m$ matrix $I_{n\times m(ij)} = \delta_{ij}$; also, $I\equiv I_{3\times 3}$. 

The contact of the stance foot can be modeled as a fictitious spherical joint attaching the biped to the ground. The universal and spherical joints can be viewed as two and three hinge joints respectively, connected in series, see Fig.(\ref{modelfig}). We use $\phi$, $\theta$ and $\psi$ to denote rotation angles in the coronal, sagittal and axial ($z=0$) planes correspondingly. In phase I, hinge angles $q_t^s = [\phi_t,\theta_t]^\T$ specify the torso's orientation relative to the legs, and hinge angles $q_l^s = [\phi_l,\theta_l,\psi_l]^\T$ specify the legs' orientation relative to the ground. In phase II, the degrees $\phi_l$ and $\psi_l$ are inactive, therefore $q_l^d = [\theta_l]$, while $q_t^d = q_t^s$. The model's configuration is fully specified by the generalized coordinates $q^p = [q_l^p;q_t^p]$, where the support phase superscripts $p$ stands for `s' and `d' in the single and double support phases respectively. In the upright configuration $q_{(i\ne 3)}^s = 0$ by convention. We will also refer to $q_\phi = [\phi_l,\phi_t]^\T$, $q_\theta = [\theta_l,\theta_t]^\T$ and  $\psi_l$ subspaces as the coronal (or $\phi$), sagittal (or $\theta$) and axial (or $\psi$) sectors correspondingly.

It is often convenient to work in a body-fixed frame of a rigid body. Let us denote the ground frame, the legs' body frame and the torso's body frame as $F_g$, $F_l$ and $F_t$ correspondingly. The ground frame $F_g$ is shown in Fig.(\ref{modelfig}). By convention, in the upright configuration $F_g = F_l = F_t$. In general, for vectors and matrices we may optionally use the frame superscript ($g$, $l$ or $t$) to explicitly indicate the frame their components are written in. For many quantities we can omit the frame superscript (unless expressly stated otherwise) when we are not concerned with a coordinate representation, or when the ground frame is implied. The rotation matrix $R_{ab}$ rotates vector coordinates from $F_a$ to $F_b$, that is $r^a = R_{ab} r^b$ for a vector $r$. The angular velocity of $F_b$ relative to $F_a$ is $\hat\Omega_{ab} = \dot R_{ab} R_{ab}^\T$, where the hat notation indicates a skew symmetric matrix $\hat a$ defined via the antisymmetric symbol as $\hat a_{(ij)} = \sum_{k}\epsilon_{(jik)}a_{(k)}$. For convenience, we may introduce shorthand notations. For example, we define $R_1 \equiv R_{gl}$, $R_t \equiv R_{lt}$, $\Omega_1 \equiv \Omega_{gl}$, $\Omega_t \equiv \Omega_{lt}$ and $\Omega_2 \equiv \Omega_{gt} = \Omega_1 + \Omega_t$. The precise specifications of the joints' parameterization, and hence the dependence of $R$ and $\Omega$ on $q$ and $\dot q$ are presented in App. \ref{rotmatparam}.

The model's geometric and inertial parameters are specified as follows. 
Let $O_0, O_1, O_2$ and $O_h$ be the foot, the legs' center of mass (COM), the torso's COM and the hip respectively, as shown in Fig.(\ref{modelfig}). We define $r_1=\overrightarrow{O_0 O_1}$, $r_2=\overrightarrow{O_0 O_2}$, $r_h=\overrightarrow{O_0 O_h}$, and $r_t=\overrightarrow{O_h O_2}$. Note, $r_1^l$, $r_h^l$ and $r_t^t$ are constants, defined by the following parameterization:
\begin{equation}
  r_1^l = 
  \begin{bmatrix} 
    0 \\ d \\ l_1 
  \end{bmatrix}
  , \quad
  r_h^l = 
  \begin{bmatrix}
    0 \\ d \\ l_h 
  \end{bmatrix}
  , \quad
  r_t^t = 
  \begin{bmatrix} 
    0 \\ 0 \\ -l_t 
  \end{bmatrix}
  .
  \label{modelrs}
\end{equation}
At the upright configuration, $r_2 = r_h + r_t = [0, d, l_2]$, where $l_2 = l_h - l_t$. In many cases, such as for COM's location $r_i$, mass $m_i$ and the moment of inertia $I_i$, we use $i=1$ and $i=2$ to designate the parameters of the legs and torso respectively. The stated model symmetries (reflections in the sagittal and coronal planes) imply that $I_i$ are diagonal in the respective body frames: $I_1^l={\rm{diag}}(I_{1\phi},I_{1\theta},I_{1\psi})$ and $I_2^t={\rm{diag}}(I_{2\phi},I_{2\theta},I_{2\psi})$.

Let us define a mass distribution moment function
\begin{equation}
  \mu_{na} = m_1 l_1^n + m_2 l_a^n
  .
\end{equation}
We will often be using the following shorthand notations:  $\mu_n \equiv \mu_{n2}$, $\tilde\mu_n \equiv \mu_{nh}$ and $I_\alpha \equiv I_{1\alpha}+I_{2\alpha}$.

\section{Equations of motion}
\label{eqmots}

In this section we first derive the exact equations of motion. These equations can only be integrated numerically. We next derive the SML equations -- an approximation to the exact equations -- that we analyze and solve analytically later on in Sec. \ref{smlsolution}. When solving numerically for a general collisionless solution, we will use the SML solution as an initial guess, which can be very helpful in a numerical solution search.

\subsection{Exact equations}
\label{exacteqmot}

The kinetic and potential energies, written in terms of the model part coordinates and velocities, are:
\begin{equation}
  \begin{split}
    & T=\frac{1}{2}\sum_{i=1,2}\left(m_i v_i^\T v_i + \Omega_i^\T I_i \Omega_i \right),
    \\
    & V = -g_f^\T \sum_{i=1,2} m_i r_i ,
  \end{split}
  \label{kinpotenergy}
\end{equation}
where: $m_i$ are masses, $I_i$ are moments of inertia, $r_i$ are COM coordinates, $v_i$ are COM velocities, $\Omega_i$ are angular velocities and $g_f = [0,0,-g]^\T$ is the gravitational field vector. The legs and torso parameters are indicated by $i = 1$ and $i = 2$ correspondingly. Following the Lagrangian formalism to derive the equations of motion \cite{landau2000mechanics,murray1994mathematical}, we need to express $T$ and $V$ in terms of the generalized coordinates $q$ and velocities $\dot q$, define the Lagrangian function $\mathcal{L}(q,\dot q) = T(q,\dot q) - V(q)$, and write down the Euler-Lagrange equations:
\begin{equation}
  \frac{d}{dt}\frac{\partial \mathcal{L}}{\partial {\dot q}} - \frac{\partial \mathcal{L}}{\partial q} = 0
  .
\end{equation}

We first consider phase I, so we assume below $q = q^s$. The kinetic energy can be written in the form $T = \frac{1}{2}{\dot q}^\T H \dot q$, where $H$ is called the mass matrix. It can be written as
\begin{equation}
  H = {\bar S}^\T \bar H \bar S,
  \label{massmatrix}
\end{equation}
where $\bar S$ is constructed from the joint rotations $R_1$ and $R_t$ (and corresponding partial rotations), and $\bar H$ is expressed in terms of the model constants ($r_1^l$, $r_h^l$, $r_t^t$, $m_1$, $m_2$, $I_1^l$, $I_2^t$) and $R_t$, see App. \ref{exacteqmotderivation} for details. Specifically, for $\bar S$ we have
\begin{equation}
  \bar S =
  \begin{bmatrix}
    R_1^\T S_1 & 0 \\
    0 & R_t^\T S_t
  \end{bmatrix}
  =
  \begin{bmatrix}
    S_1^l & 0 \\
    0 & S_t^t
  \end{bmatrix}
  ,
  \label{bars}
\end{equation}
where
\begin{equation}
  S_1 =
  \begin{bmatrix}
    \left(R_Z(\psi_l)R_Y(\theta_l)\right)_{(\cdot 1)}, 
    R_Z(\psi_l)_{(\cdot 2)}, 
    I_{(\cdot 3)}
  \end{bmatrix}
  \label{s1}
\end{equation}
and
\begin{equation}
  S_t =
  \begin{bmatrix}
    R_Y(\theta_t)_{(\cdot 1)}, 
    I_{(\cdot 2)}
  \end{bmatrix}
  .
  \label{st}
\end{equation}
For $\bar H$ we have
\begin{equation}
  \bar H = 
  \begin{bmatrix}
    \bar H_{ll} & \bar H_{lt} \\
    \bar H_{tl} & \bar H_{tt}
  \end{bmatrix}
  ,
\end{equation}
where 
\begin{equation}
  \begin{split}
    & \bar H_{ll} = m_1 {\widehat {r_1^l}}^\T \widehat {r_1^l} + I_1^l + R_t \left( m_2 {\widehat{r_2^t}}^\T \widehat{r_2^t} + I_2^t \right) R_t^\T , \\
    & \bar H_{lt} = {\bar{H}_{tl}}^\T = R_t \left( m_2{\widehat{r_2^t}}^\T \widehat {r_t^t} + I_2^t \right) , \\
    &  \bar H_{tt} = m_2 {\widehat {r_t^t}}^\T \widehat {r_t^t} + I_2^t ,
    \label{barh}
  \end{split}
\end{equation}
where $\widehat{r_2^t} = \widehat{R_t^\T r_h^l} +\widehat{r_t^t}$.

Following the standard procedure \cite{murray1994mathematical} we use the mass matrix $H$ to define the Coriolis matrix $C$ in terms of the Christoffel symbols $\Gamma_{(ijk)}$ as 
\begin{equation}
  C=\sum_k \Gamma^k \dot q_{(k)},
  \label{coriolis}
\end{equation}
where
\begin{equation}
  \Gamma^k_{(ij)} = \Gamma_{(ijk)} = \frac{1}{2}\left(\frac{\partial H_{(ij)}}{\partial q_{(k)}} + \frac{\partial H_{(ik)}}{\partial q_{(j)}} - \frac{\partial H_{(jk)}}{\partial q_{(i)}} \right)
  \label{christoffel}
  .
\end{equation}

The potential energy, written explicitly in terms of the model constants and joint rotations, is:
\begin{equation}
  V = g(R_1(m_1 r_1^l + m_2 (r_h^l + R_t r_t^t)))_{(3)}
  \label{potenviarelrots}
\end{equation}
where $g$ is the gravity acceleration magnitude. The potential term $G$ is defined as the gradient of the potential energy $V(q)$ 
\begin{equation}
  G = \nabla V.
  \label{potterm}
\end{equation}

We are now in the position to write down the equations of motion in the manipulator equation form \cite{murray1994mathematical}:
\begin{equation}
  H \ddot q + C \dot q + G = 0
  \label{eqmotion}
\end{equation}
where $H$, $C$ and $G$ have been defined above. These terms are defined explicitly in terms of the model parameters, save for differentiation with respect to $q$, which can be done either symbolically (manually or with help of differentiation software) or numerically. The equation can then be integrated numerically.

We next consider phase II. The legs coordinates $q_l$ change from $[\phi_l, \theta_l, \psi_l]^\T$ in phase I to $[\theta_l]$ in phase II. Correspondingly, $R_1$ becomes $R_Y(\theta_l)$ and $S_1$ becomes $I_{(\cdot 2)}$. The equations of motion in phase II can be formally written as a projection of the phase I equations. To that end we introduce a projection operator $J^d$ that projects $q^s$ onto $q^d$: $q^d = J^d q^s$. It is defined as $J^d_{(1,2)} = J^d_{(2,4)} = J^d_{(3,5)} = 1$ with all other entrees equal to zero. We also define a projection operator $J^s = {J^d}^\T J^d$, which zeros out $q^s$ components that are absent in $q^d$, namely $q^s_{(1)}$ and $q^s_{(3)}$. The phase I Eq.(\ref{eqmotion}) can be written as $D^sq = 0$, where $D^s(q,\dot q) = H(q) \frac{d^2}{dt^2} + C(q,\dot q) \frac{d}{dt} + \lVert q\rVert^{-2}G(q)q^\T$. The phase II equations of motion $D^d q^d = 0$ can then be obtained by projecting:
\begin{equation}
  D^d = J^d D^s(J^s q, J^s \dot q) {J^d}^\T
  \label{dprojection}
  .
\end{equation}
We next consider the equations of motion in the SML approximation.

\subsection{Small movement limit}
\label{sml}

The small movement limit is essentially a relationship between relevant problem scales in the limit of small feet separation that justifies perturbation theory treatment in which the dynamics of the $\psi$ sector are decided at higher order by the (decoupled at lower order) $\phi$ and $\theta$ sectors. To specify it precisely, we introduce dimensionless parameters: dimensionless foot separation parameter $\bar d = \mu_0d/\mu_1$ and dimensionless axial moment of inertia $\bar I_\psi = I_\psi/\mu_2$. We will see in this section that $\phi \sim \bar d$, $\theta \sim \varepsilon$ and $\psi \sim \bar d \varepsilon / \bar I_\psi$, where $\varepsilon$ is a free scale parameter, independent of the model parameterization. For the consistency of our perturbative analysis, we need $1 \gg \phi, \theta \gg \psi, \phi^2, \theta^2$, from where we obtain the SML specifications:
\begin{equation}
  {\rm SML:} \quad 
  \left\{
  \begin{split}
    \bar{d} \ll 1 ,\\
    {\bar{d}}^2 \ll \varepsilon \ll \sqrt{\bar{d}}, \\
    \bar{I}_{\psi} \gg \max{\{\bar{d}, \varepsilon \}}.
  \end{split}
  \right.
  \label{smlspecs}
\end{equation}
A more careful analysis reveals that these conditions can be relaxed somewhat (e.g. the third line can be replaced by $\bar{I}_{\psi} \gg \bar{d}^2$), but for the sake of simplicity we define the SML as above. Below we expand the exact equations of motion to the lowest order sufficient for establishing a nontrivial walking solution.

First we consider phase I.
In the linear dynamics limit the equations of motion are
\begin{equation}
  H_0 \ddot q + G_0 + G_1 q = 0,
  \label{lindyneqmot}
\end{equation}
where $H_0$, $G_0$ and $G_1$ are derived in App. \ref{lindynapprox}, (nonperturbatively in $\bar d$, i.e. correct to all orders in $\bar d$).
In this limit, the $\phi$ sector is completely decoupled from the other sectors. The $\theta$ sector is coupled to the $\psi$ sector via $\mathcal{O}(\bar d)$ coupling. To the lowest nontrivial order in $\bar d$, (namely to order $\mathcal{O}(1)$ for $H_0$ and $G_1$, and to order $\mathcal{O}(\bar d)$ for $G_0$, which coincidentally are exact expressions for $G_0$ and $G_1$), the equations of motion in $\alpha \in \{\phi, \theta\}$ sectors assume a similar form:
\begin{equation}
  \begin{split}
    & H_\alpha = 
    \begin{bmatrix}
      \mu_2 + I_{\alpha} & -m_2 l_2 l_t + I_{2\alpha} \\
      -m_2 l_2 l_t + I_{2\alpha} & m_2 l_t^2 + I_{2\alpha}
    \end{bmatrix}
    , \\
    & G_{0\alpha} = g
    \begin{bmatrix}
      \delta_{\alpha\phi} \mu_0 d \\
      0
    \end{bmatrix}
    , \quad
    G_{1\alpha} = g 
    \begin{bmatrix}
      -\mu_1 &  m_2 l_t \\
      m_2 l_t & m_2 l_t
    \end{bmatrix}
    .
  \end{split}
  \label{hg0g1alpha}
\end{equation}
Note, $G_{0\phi}$ is $\mathcal{O}(\bar d)$, while $G_{0\theta} = 0$. As a consequence, the scale of $\phi$ sector motion is controlled by $\bar d$, while the scale $\varepsilon$ of $\theta$ sector motion is not determined by the model parameters, but only needs to be $\varepsilon \ll 1$ to justify the linear dynamics approximation.

To the considered order (as in Eq.(\ref{hg0g1alpha})), $\ddot\psi_l = 0$ and the solution describes a walking-in-place motion (walking with zero step size). To capture a non-trivial motion in $\psi$ sector, we need to include the terms to order $\mathcal{O}(\bar d \varepsilon)$ in the equations of motion, for the purpose of computing $\psi_l$. This entails including the terms of order $\mathcal{O}(\bar d)$ from $H_0$, as well as the order $\mathcal{O}(q)$ corrections to the mass matrix $H$ and Coriolis term $C$, see App. \ref{nextorder} for details. The equations of motion for $\psi_l$ in the small movement limit become:
\begin{equation}
  I_\psi \ddot\psi_l + d b^\T \ddot q_\theta + 
  \frac{d}{dt}\left( q_\phi^\T A_{\phi\theta} \dot q_\theta +
  q_\theta^\T A_{\theta\phi} \dot q_\phi \right) = 0,
  \label{eqmotpsi}
\end{equation}
where
\begin{equation}
  b = 
  \begin{bmatrix}
    -\mu_1 \\
    m_2 l_t
  \end{bmatrix}
  ,
\end{equation}
and
\begin{equation}
  \begin{split}
    & A_{\phi\theta} = 
    \begin{bmatrix}
      \mu_2 + I_\theta - I_\psi & -m_2 l_2 l_t + I_{2\theta}  \\
      -m_2 l_2 l_t + I_{2\theta} - I_{2\psi} & m_2 l_t^2 +I_{2\theta}- I_{2\psi}
    \end{bmatrix}
    , \\
    & A_{\theta\phi} = 
    \begin{bmatrix}
      -\mu_2 - I_\phi & m_2 l_2 l_t - I_{2\phi}  \\
      m_2 l_2 l_t - I_{2\phi} + I_{2\psi} & -m_2 l_t^2
    \end{bmatrix}.
  \end{split}
  \label{aptatp}
\end{equation}
These equations properly account for all the dominant terms up to $\mathcal{O}(\bar d \varepsilon)$, as long as the SML conditions of Eq.(\ref{smlspecs}) are satisfied.

In the small movement limit, rather than working with relative angles (joint angles) as generalized coordinates in $\alpha \in \{\phi, \theta\}$ sectors, we found it convenient to work with absolute angles measured with respect to the statically balanced single support configuration. ($G$ becomes diagonal in that basis). Therefore, we switch to new coordinates $\tilde q_\alpha = [\tilde\alpha_l, \tilde\alpha_t]^\T$, defined as:
\begin{equation}
  \tilde q_\alpha = \tilde S q_\alpha + c_\alpha,
\end{equation}
where $\tilde S = [1,0;1,1]$ and vector $c_\alpha$ is chosen to cancel the constant term $G_{0\alpha}$. Let us write the mass matrix $H_\alpha$ and the potential term $G_{1\alpha}$ as
\begin{equation}
  H_\alpha = \tilde S^\T \tilde H_\alpha \tilde S, \quad
  G_{1\alpha} = \tilde S^\T \tilde G_{1\alpha} \tilde S.
  \label{tHatG1a}
\end{equation}
Then the equations of motion in $\alpha\in\{\phi,\theta\}$ sectors become
\begin{equation}
  \tilde H_\alpha \ddot{\tilde q}_\alpha + \tilde G_{1\alpha} \tilde q_\alpha = 0
  \label{eqmotnatural}
\end{equation}
where $\tilde H_\alpha$ and $\tilde G_{1\alpha}$ are
\begin{equation}
  \begin{split}
    & \tilde H_\alpha = 
    \begin{bmatrix}
      \tilde\mu_2 + I_{1\alpha} & -m_2 l_h l_t \\
      -m_2 l_h l_t & m_2 l_t^2 + I_{2\alpha}
    \end{bmatrix}
    , \\
    & \tilde G_{1\alpha} = g 
    \begin{bmatrix}
      -\tilde\mu_1 & 0 \\
      0 & m_2 l_t
    \end{bmatrix}
  \end{split}
  \label{hg1tilde}
\end{equation}
and the term $c_\alpha$ was set to $(\tilde S^\T \tilde G_{1\alpha})^{-1} G_{0\alpha}$:
\begin{equation}
  c_\alpha = 
  \begin{bmatrix}
    -\delta_{\alpha\phi}\frac{\mu_0 d}{\tilde\mu_1} \\
    0
  \end{bmatrix}
  .
\end{equation}
Note that $\tilde G_{1\alpha}$ (and $G_{1\alpha}$) does not depend on $\alpha$, but we keep the subscript to distinguish it from $G_1$. 

Projecting the equations of motion onto phase II, (see Eq.(\ref{dprojection})), we find that the $\phi$ sector reduces to $\phi_t$ governed by the equation
\begin{equation}
  (m_2 l_t^2 + I_{2\phi}) \ddot\phi_t +g m_2 l_t \phi_t = 0
  \label{eqmotphip2}
  ,
\end{equation}
the $\theta$ sector equations remain unchanged, and the $\psi$ sector is absent in phase II.

\section{Solution symmetries and constraints}
\label{solutionconstrs}

Similarly to Ref. \cite{gomes2011walking}, we will impose certain symmetry constraints on the form of the solution of Eq.(\ref{eqmotion}). This will reduce the number of equations we need to consider and will make the problem more manageable, which is especially important for an analytical investigation.

The model Lagrangian, and therefore the derived equations of motion, are invariant with respect to time reversal and a spatial reflection across a vertical plane. Therefore, a collisionless solution remains valid under these transformations. Due to the model's sagittal and coronal plane symmetries, we can look for a solution that is invariant to simultaneous time reversal and spatial reflections. Specifically, we require the invariance of the solution in phase I under $(t,x) \to -(t,x)$, and in phase II under $(t,x,y) \to -(t,x,y)$, (where $(t,x,y)$ are measured relative to certain symmetry points). We call these symmetry points $P_s$ and $P_d$, for the single and double support phases respectively. In terms of the generalized coordinates, the solution is invariant under the following transformations:
\begin{equation}
  \begin{split}
    & P_s: \quad (t,\theta_l,\psi_l,\theta_t) \to -(t,\theta_l,\psi_l,\theta_t) \\
    & P_d: \quad (t,\theta_l,\phi_t,\theta_t) \to -(t,\theta_l,\phi_t,\theta_t)
  \end{split}
  \label{symmetries}
\end{equation}

The generalized coordinates and their time derivatives are continuous functions of time at the symmetry points. Therefore, if a coordinate flips sign in Eq.(\ref{symmetries}), the coordinate and its even order time derivatives must vanish at the symmetry point. Otherwise, the odd order time derivatives vanish. Notice also that the symmetry of $P_d$ implies that the walker is in the upright configuration at $P_d$.

Thanks to the symmetries of Eq.(\ref{symmetries}), the complete walking cycle can be obtained from the part of the trajectory connecting nearby points $P_s$ and $P_d$, that makes up a quarter of the cycle. Thus, it is sufficient to focus on the solution between the two symmetry points.

 The transition from phase I to phase II is punctuated by a foot impact event. In a time-reversed picture, in the phase II to phase I transition, this event would normally be viewed as a foot lift-off. We will refer to it as an impact in either case.

Let $q^s(t)$ be the result of integrating the equations of motion $D^sq^s = 0$ forward in time starting from $P_s$. Let $q^d(t)$ be the result of integrating the equations of motion $D^dq^d = 0$ backward in time starting from $P_d$. Let $t_s$ and $-t_d$ be the moment of time $t$ where the respective branches reach the impact point. For a collisionless impact it is required \cite{chatterjee2002persistent} that the swing foot position $r$ (of the branch $q^s(t)$) satisfy at the impact $\dot r = \ddot r =0$. Since the feet positions are not affected by $\theta_l$ rotations, and $\phi_l=0$ in phase II, the impact conditions translate into $\phi_l^s=\dot\phi_l^s = \dot\psi_l^s = \ddot\phi_l^s = \ddot\psi_l^s = 0$ at $t=t_s$. We can write the impact conditions as
\begin{equation}
  D^i q^s(t_s) = 0,
\label{impactconds}
\end{equation}
where $D^i_{(1,1)}=1$, $D^i_{(2,1)}=D^i_{(3,3)}=\frac{d}{dt}$, $D^i_{(4,1)}=D^i_{(5,3)}=\frac{d^2}{dt^2}$, and the rest are zeros. For $q^s(t)$ and $q^d(-t)$ to belong to the same solution, the model states from two branches must match at the impact: $J^d q^s(t_s) = q^d(-t_d)$ and $J^d \dot q^s(t_s) = \dot q^d(-t_d)$. We can write the matching conditions as
\begin{equation}
  D^m \left(J^d q^s(t_s) - q^d(-t_d)\right) = 0
\label{matchconds}
\end{equation}
where $D^m = [I;I \frac{d}{dt}]$.

To find a valid collisionless solution, there are 11 constraints to be satisfied: 5 from Eq.(\ref{impactconds}) and 6 from Eq.(\ref{matchconds}). We will refer to them as joining conditions, as they are imposed at the joining point of two branches. If the model is fixed, the number of free parameters that can be varied is 10: 8 initial conditions on the parameters that are invariant to the symmetry transformations Eq.(\ref{symmetries}), namely $(\phi_l,\dot\theta_l,\dot\psi_l,\phi_t,\dot\theta_t)|_{t=0}$ for $P_s$ and $(\dot\theta_l,\dot\phi_t,\dot\theta_t)|_{t=0}$ for $P_d$, and 2 impact times, $t^s$ and $t^d$. Even if all the parameters are independent (which is not the case in the linear dynamics limit, as we show later), it is not possible to construct a lossless gait without tuning the model parameters.

We can generalize this consideration to other models with periodic collisionless solutions \cite{chatterjee2002persistent,gomes2005collisionless,gomes2011walking}. In general, let $P_s$ and $P_d$ be the symmetry points in the less constrained (phase I) and more constrained (phase II) phases respectively. In a symmetry point, for every degree of freedom, either its coordinate or velocity turns zero. A collisionless solution is invariant with respect to a sign flip in time and in every component turning zero. Let $n_s = \dim(q^s)$ and $n_d = \dim(q^d)$. We define the impact dimension $d_i$ as the number of degrees that freeze (become inactive, in other words) upon transitioning from phase I to phase II: $d_i = n_s - n_d$. Let $n_e$ be the number of parameters encoding the impact surface. In general, $ n_e = \dim(space)-\dim(ground)$, so $n_e = 1$ in all the cited cases. The number of adjustable solution parameters is $n_s + n_d + 2$: one from each degree of freedom (non-zero coordinate or velocity) plus the impact time, for both phases. The number of joining conditions is $n_e + 2d_i + 2n_d$: $n_e + 2d_i$ impact conditions and $2n_d$ matching conditions. Thus, the number of model parameters requiring tuning is $n = (n_e + 2d_i + 2n_d) - (n_s + n_d + 2) = d_i + n_e - 2$. Note that in general $n_e$ can differ from 1; for example, $n_e = 2$ for a three-dimensional tightrope walker.

\section{Small movement limit solution}
\label{smlsolution}

In the SML, the equations in the $\phi$ and $\theta$ sectors are linear. While the equation on $\psi$ is nonlinear, it can be easily integrated analytically. However, the need to coordinate motion in three planes to satisfy the joining conditions of a collisionless solution leads to nontrivial nonlinear equations on the impact times and normal mode frequencies. These equations are difficult to analyze in their generality. In this section, the analysis is simplified by imposition of an additional constraint and a formulation in terms of dimensionless impact phases. Also, we will focus our attention on the least exotic walking solution, that is a solution with the fewest oscillations.

\subsection{Axial plane solution}

To distinguish quantities in phases I and II, we optionally use the phase notation $p\in\{s,d\}$ that assumes the value `s' in phase I (single support) and the value `d' in phase II (double support). To reduce clutter we omit the sector $\psi$ and phase $s$ superscripts in this section, listing all the affected quantities at the end.

Integrating Eq.(\ref{eqmotpsi}) we find
\begin{align}
  \psi_l(t) & = w t - I_\psi^{-1} \biggl( db^\T q_\theta(t) \\
  & + \int_0^t dt' \left( q_\phi^\T(t') A_{\phi\theta} \dot q_\theta(t') 
  + q_\theta^\T(t') A_{\theta\phi} \dot q_\phi(t') \right) \biggr)
  \label{psisol}
  \notag
  .
\end{align}
Note, a constant term in the above expression is absent due to the symmetry of $P_s$ (see Eq.(\ref{symmetries})), dictating that $\psi_l(t)$ and $q_{\theta}(t)$ be odd functions, and $q_{\phi}(t)$ to be an even function.

To significantly simplify our analysis of the SML solution, we will impose an additional constraint on it, requiring the vanishing of $q_\theta$ at the impact: 
\begin{equation}
  q_\theta(t_s) = 0.
  \label{uprightlegs}
\end{equation}
We call it the upright legs at impact (ULI) constraint, as it implies the same configuration of the legs, as in the upright configuration. As will be explained in Sec. \ref{sagplanesol}, in the SML, $q_\theta(t)$ realizes a simple harmonic motion, and therefore, the ULI constraint also implies $\ddot q_\theta(t_s) = 0$. To satisfy the constraint we will need to tune an additional model parameter. The constraint is imposed only in the SML, and not on the general solution.

There are two joining equations involving $\psi_l(t)$:
\begin{equation}
  \begin{split}
    & \dot\psi_l(t_s) = 0, \\
    & \ddot \psi_l(t_s) = 0.
  \end{split}
  \label{axialjoinconds}
\end{equation}
Both equations arise from the impact conditions Eq.(\ref{impactconds}). Under the ULI constraint, the joining conditions in Eq.(\ref{axialjoinconds}) translate respectively to
\begin{equation}
  \begin{split}
    & w - I_\psi^{-1} \left( db^\T 
    + q_\phi^\T(t_s) A_{\phi\theta} \right) \dot q_\theta(t_s) = 0, \\
    & \dot q_\phi^\T(t_s) \left( A_{\phi\theta} 
    + A_{\theta\phi}^\T \right) \dot q_\theta(t_s) = 0.
  \end{split}
  \label{axialjoinconds1}
\end{equation}
At the impact point, $\dot \phi_l(t_s) = 0$, see Eq.(\ref{impactconds}). This also implies $\dot \phi_t(t_s) \ne 0$, because otherwise $[q_\phi(t_s);\dot q_\phi(t_s)]$ would be a singular point on the phase portrait of the $\phi$ sector dynamics (which is decoupled from other sectors in the SML), which would be incompatible with a periodic walking solution we are looking for. Therefore, the second joining condition becomes: 
\begin{equation}
    b_1^\T \dot{\tilde q}_\theta(t_s) = 0,
  \label{axialjoincond2}
\end{equation}
where $b_1 = \left(\tilde S^{-1}\right)^\T \left( A_{\phi\theta}^\T + A_{\theta\phi} \right)_{(\cdot 2)} = [-I_{2\phi}; I_{2\theta}-I_{2\psi}]$.

In this section, we have omitted $\psi$ and $s$ superscripts in: $\psi_l^s$, $w^{\psi s}$, $q_\theta^s$, $q_\phi^s$, $\phi_l^s$ and $\phi_t^s$.

\subsection{General form of linear dynamics solution in sagittal and coronal sectors}

Let us investigate the equation Eq.(\ref{eqmotnatural}) that describes motion in both the coronal ($\alpha=\phi$) plane in phase I, and sagittal ($\alpha=\theta$) plane in phases I and II. To reduce notational clutter, we will be omitting the sector $\alpha$ and phase $p$ sub- and superscripts throughout this section. To eliminate ambiguity, we provide a list of affected quantities in full notation at the end of the section. Let us introduce a matrix $M$:
\begin{equation}
  M = - \tilde G_1^{-1} \tilde H =
  \begin{bmatrix}
    a_+ & -\gamma\beta^{-1} \\
    \gamma\beta & a_-
  \end{bmatrix}
  \label{Mmat}
  ,
\end{equation}
where
\begin{equation}
  \begin{split}
    & a_+ = \frac{\tilde\mu_2 + I_1}{g\tilde\mu_1}, \quad
    a_- = - \frac{m_2 l_t^2 + I_2}{g m_2 l_t}, \\ 
    & \beta = \sqrt{\frac{\tilde\mu_1}{m_2 l_t}}, \quad
    \gamma = \frac{l_h}{g\beta}.
  \end{split}
  \label{azbetagamma}
\end{equation}
Let $\{\lambda_z,u_z\}|_{z=\pm}$ be the eigensystem of $M$. The subscript $z$ represents $\pm 1$, which we often shorten to $\pm$. Notice that $\det{M} < 0$, because $\tilde H$ is positive definite and $\det{\tilde G_1} < 0$, see Eq.(\ref{hg1tilde}). Therefore, $M$ has two real eigenvalues, one positive and one negative (see App. \ref{miscrels} for details): $\lambda_+ > 0$ and $\lambda_- < 0$. The general solution of Eq.(\ref{eqmotnatural}) can be written as
\begin{equation}
  \tilde q(t)=\sum_{zz'=\pm1}w_{zz'} u_z e^{\frac{z't}{\sqrt{\lambda_z}}}
  \label{qgennatural}
  ,
\end{equation}
where $w_{zz'}$ are the weights of different normal modes. Let us define a function
\begin{equation}
  \Lambda_{zz'}(a,b,c)=\frac{1}{2}\left(a+zb+z'\sqrt{(a-b)^2+4c} \right)
  \label{Lambda}
\end{equation}
and shorthand notations
\begin{equation}
  \begin{split}
    & \lambda_{zz'}\equiv \Lambda_{zz'}(a_+, a_-, -\gamma^2),
    \\
    & a_{zz'}\equiv \Lambda_{zz'}(\lambda_+,\lambda_-,\gamma^2).
  \end{split}
  \label{lambdazzazz}
\end{equation}
We can now compactly write down the eigensystem of $M$
\begin{equation}
  \lambda_z=\lambda_{+,z},
  \quad
  u_z=\left[\lambda_{-,z}, \gamma\beta \right]^\T
  \label{eigensystem}
\end{equation}
and the expression of $a_z$ in terms of $\lambda_z$ and $\gamma$
\begin{equation}
  a_z = a_{+,z}
  \label{azfromlambdaz}
  .
\end{equation}

The general solution of the phase II equation Eq.(\ref{eqmotphip2}) is
\begin{equation}
  \phi_t(t)=\sum_{z=\pm1}w_{z} e^{\frac{zt}{\sqrt{a_-}}}
  \label{phigenp2}
  .
\end{equation}

The quantities in this section, that have the sector $\alpha$ and phase $p$ sub- or superscript in their full notations, include: $M_{\alpha}$, $\tilde G_{1\alpha}$, $\tilde H_\alpha$, $a_z^\alpha$, $I_{i\alpha}$, $\lambda_z^{\alpha p}$, $u_z^{\alpha p}$, $\tilde q_\alpha^p$, $w_{zz'}^{\alpha p}$, $\lambda_{zz'}^{\alpha p}$ and $a_{zz'}^{\alpha p}$.

We will be using a frequency notation $\omega_z^{\alpha p} \equiv 1/\sqrt{z\lambda_z^{\alpha p}}$ in the following analysis. If a quantity does not depend on the phase index $p$, as is the case for $\lambda_z^{\theta p}$, $u_z^{\theta p}$ and $\omega_z^{\theta p}$, we will omit $p$ for clarity.

\subsection{Sagittal plane solution}
\label{sagplanesol}

To reduce clutter we omit the sector $\theta$ superscript in this section, listing the affected quantities at the end.

According to Eq.(\ref{symmetries}), the symmetries of $P_s$ and $P_d$ dictate that $\tilde q_{\theta}^p(t)$ be an odd function in both phases. Therefore, the weights $w_{zz'}^{p}$ in Eq.(\ref{qgennatural}) must be odd functions of $z'$. Consequently, the sagittal plane solution must be in the form:
\begin{equation}
  \tilde q_\theta^p(t) = w_+^{p} u_+ \sinh{\omega_+ t} + w_-^{p} u_- \sin{\omega_- t}
  ,
  \label{thetasol}
\end{equation}
where we have defined $w_z^{p} = 2z\sqrt{z}z'w_{zz'}^{p}$. 
There are four scalar joining equations involving the sagittal sector:
\begin{equation}
  \begin{split}
    & {\tilde q}_\theta^s(t_s) - {\tilde q}_\theta^d(-t_d) = 0, \\
    & \dot {\tilde q}_\theta^s(t_s) - \dot {\tilde q}_\theta^d(-t_d) = 0 .
  \end{split}
  \label{sagittaljoinconds}
\end{equation}
All the equations arise from the matching conditions Eq.(\ref{matchconds}). Because the equations of motion are identical in both phases, a formal continuation of either branch $\tilde q_\theta^p(t)$ to all $t$ is a valid solution (up to a time shift), and
\begin{equation}
  \tilde q_\theta^s(t) = \tilde q_\theta^d(t-t_s-t_d).
  \label{qthetasdrel}
\end{equation}
Since the solution is periodic and the first term in the right hand side of Eq.(\ref{thetasol}) is unbounded, its weight must be $w_+ = 0$ and the sagittal sector motion in both phases is a simple harmonic motion $\tilde q_\theta^p(t) \propto \sin{\omega_- t}$. It then follows from Eq.(\ref{qthetasdrel}):
\begin{equation}
  \omega_- (t_s + t_d) = \pi k, \,\, k\in\mathbb{Z}
  \label{tsplustd}
\end{equation}
(where $\mathbb{Z}$ is the set of integer numbers) and $w_-^{s} = (-1)^k w_-^{d}$.
The ULI constraint of Eq.(\ref{uprightlegs}) implies $\omega_- t_s = \pi k'$, $k'\in\mathbb{Z}$, and therefore $\omega_- t_d = \pi (k-k')$. In the rest of the analysis, we choose to consider a solution with the smallest positive $t_p$, therefore we pick $k = 2, k' = 1$ to have:
\begin{equation}
  t_s = t_d = \frac{\pi}{\omega_-}.
  \label{tstdsagittal}
\end{equation}
Because we are focusing on a solution with $t_s = t_d$, we may use a single notation $\tau = t_s = t_d$ to represent the impact times in both phases. We introduce an impact phase notation $o_- = \omega_- \tau$, so the previous equation can be written as
\begin{equation}
  o_- = \pi
  \label{osagittal}
  .
\end{equation}
An impact phase describes the phase gained by a normal mode between $t=0$ and the impact moment. The full utility of the impact phase notations will become evident in the next section.

In this section, we have omitted $\theta$ superscripts in: $w_{zz'}^{\theta p}$, $w_z^{\theta p}$, $u_z^\theta$, $\omega_z^\theta$ and $o_-^\theta$.

\subsection{Coronal plane solution}
\label{corplanesol}

To reduce notational clutter we omit the sector $\phi$ superscript in this section, listing all the affected quantities at the end.

According to Eq.(\ref{symmetries}), the symmetries of $P_s$ and $P_d$ dictate that $\tilde q_{\phi}^s(t)$ and $\phi_t^d(t)$ be an even and odd functions respectively. Therefore, the weights $w_{zz'}^{s}$ in Eq.(\ref{qgennatural}) must be an even function of $z'$, while the weights $w_z^{d}$ in Eq.(\ref{phigenp2}) must be an odd function of $z$. Consequently, the coronal plane solution must have the form:
\begin{equation}
  \begin{split}
    & \tilde q_\phi^s(t) = w_+^{s} u_+^{s} \cosh{\omega_+^{s} t} + w_-^{s} u_-^{s} \cos{\omega_-^{s} t}, \\
    & \phi_t^d(t) = w^{d} \sin{\omega^{d} t},
  \end{split}
  \label{phisol}
\end{equation}
where we have defined $w_z^{s} = 2w_{zz'}^{s}$, $w^{d} = -2izw_z^{d}$ and $\omega^{d} = 1/\sqrt{-a_-}$.
There are five joining equations involving the coronal sector:
\begin{equation}
  \begin{split}
    & {\tilde \phi}_l^s(t_s) - c_{\phi(1)} = 0, \\
    & \dot{\tilde \phi}_l^s(t_s) = 0, \\
    & \ddot{\tilde \phi}_l^s(t_s) = 0, \\
    & {\tilde \phi}_t^s(t_s) - \phi_t^d(-t_d) = 0, \\
    & \dot{\tilde \phi}_t^s(t_s) - \dot\phi_t^d(-t_d) = 0.
  \end{split}
  \label{coronaljoinconds}
\end{equation}
The first three equations arise from the impact conditions Eq.(\ref{impactconds}), while two other equations arise from the matching conditions Eq.(\ref{matchconds}), (the last two equations were simplified using the first two). The five joining conditions form a set of linear equations on $w=[w_+^{s}, w_-^{s}, w^{d}]^\T$. The first equation determines the scale of the solution of the homogeneous linear system $A(t_s,t_d)w = 0$, comprised of the other four equations. The matrix $A(t_s,t_d)$ is a $4 \times 3$ matrix, whose coefficients are functions of $t_s$ and $t_d$. For a nontrivial $w$ solution to exist, the rank of $A$ must be lowered below 3. This can be achieved by appropriately tuning $t_s$ and $t_d$. To that end, we have derived the following equations on $t_p$, (see App. \ref{coronalimptimes} for details):
\begin{equation}
  \begin{split}
    & \omega_-^{s} \cot{\omega_-^{s} t_s} = \omega_+^{s} \coth{\omega_+^{s} t_s} ,\\
    & \omega_-^{s} \cot{\omega_-^{s} t_s} = \omega^{d} \tan{\omega^{d} t_d} .
  \end{split}
  \label{tstdcoronal}
\end{equation}
Because the equation of motion in phase II is a projection of the phase I equations, the frequencies $\omega_z^{s}$ and $\omega^{d}$ are not independent, they satisfy (see App. \ref{miscrels}):
\begin{equation}
  (a_- - \lambda_-)(a_- - \lambda_+) = \gamma^2
  \label{alambdagamma}
  .
\end{equation}
As was discussed in the previous section, we consider a solution with $t_s = t_d = \tau$. We can view Eqs.(\ref{tstdcoronal},\ref{alambdagamma}) as a system of equations on the $\phi$ sector spectrum, given the impact time $\tau$. We can write them as
\begin{equation}
  \begin{split}
    & o_+\coth{o_+} = o_-\cot{o_-} = o\tan{o} , \\
    & \left(o^{-2}-o_-^{-2}\right)\left(o^{-2}+o_+^{-2}\right)={\bar\tau}^{-4} ,
  \end{split}
  \label{ooo}
\end{equation}
where we have introduced the impact phase notations $o_z = \omega_z^{s} \tau$, $o=\omega^{d} \tau$ and the reduced impact time notation $\bar\tau = \tau/\sqrt{\gamma}$. Interestingly, the equations in Eq.(\ref{ooo}) are free of any model parameters. In that sense the impact phases are universal. We refer to them, and any other functions derived from the impact phases and the reduced impact time alone, as universal functions. 
\begin{figure}
  \includegraphics[width=1\columnwidth]{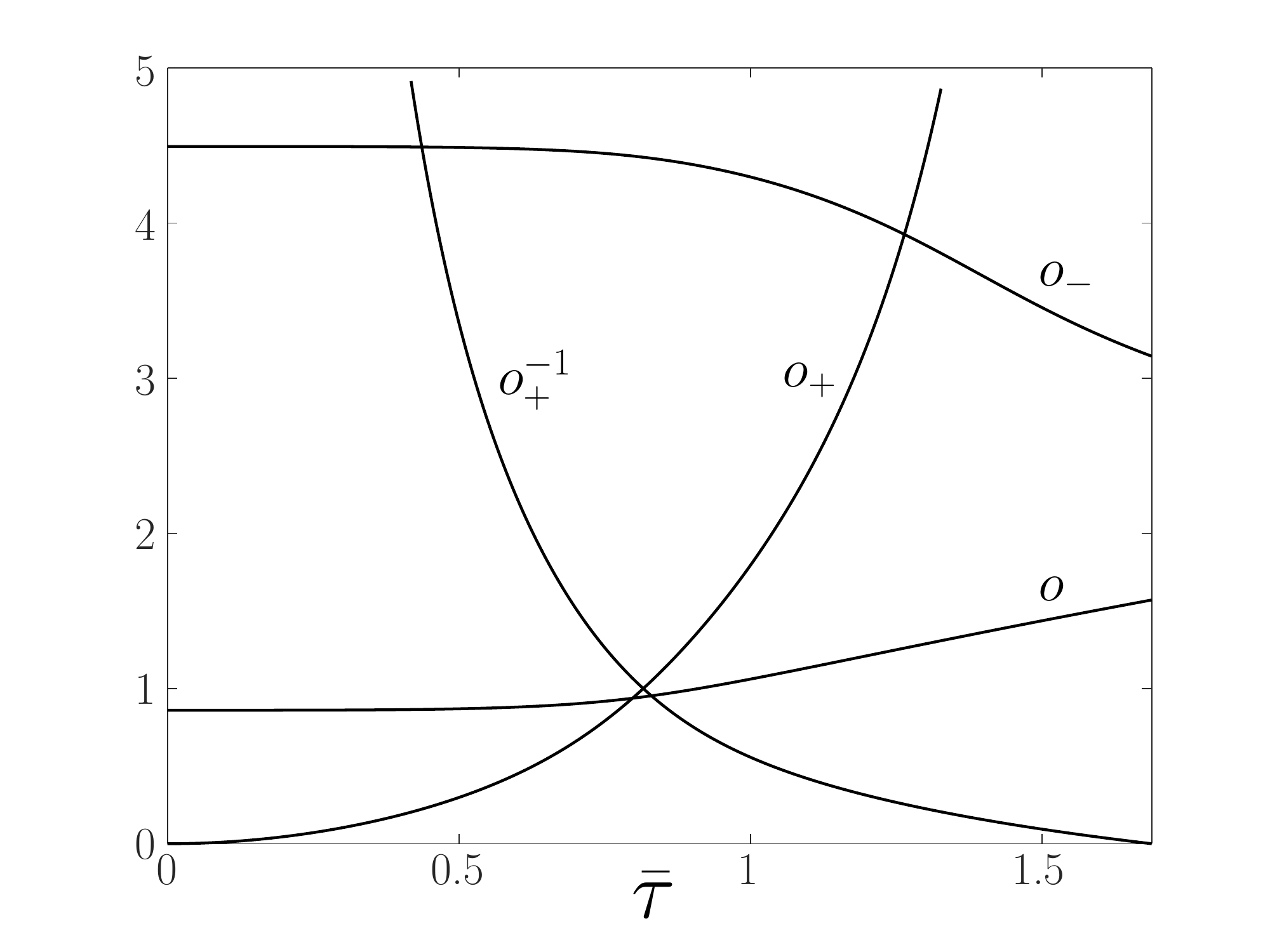}
  \caption{The impact phases $o_+(\bar\tau)$, $o_-(\bar\tau)$ and $o(\bar\tau)$ are functions of the reduced impact time $\bar\tau \in (0, \bar\tau_c)$. 
Because $o_+$ is unbounded, it is plotted together with $1/o_+$.}
  \label{ooofig}
\end{figure}

 Since $\tau$ and the frequencies are positive, we are only concerned with positive roots of Eq.(\ref{ooo}). Since $o_+\coth{o_+} \ge 1$, both $o_-$ and $o$ can only belong to $(k\pi,(k+1/2)\pi)$, $k\in\mathbb{Z}$, where $\tan{()}$ is positive. Since $o_-\cot{o_-} \le 1$ for $o_- < \pi/2$, we need $k>0$ for $o_-$. The parameter $k$ characterizes how many oscillating movements the walker commits before the impact. We choose to look for a solution with the minimal number of such oscillations, therefore we limit our consideration to $o_- \in (\pi,3\pi/2)$ and $o \in (0,\pi/2)$ in the rest of this section. Under these constraints, $o_+\coth{o_+}$, $o_-\cot{o_-}$ and $o\tan{o}$ are monotonic functions of their arguments. It then follows from the first line of Eq.(\ref{ooo}) that $o_+$ and $o$ are monotonically decreasing functions of $o_-$. From the second line of Eq.(\ref{ooo}) it follows that $\bar\tau$ is also a monotonically decreasing function of $o_-$. Therefore, all the impact phases are monotonic functions of $\bar\tau$: $o_-$ is decreasing, while $o_+$ and $o$ are increasing functions. The solution of Eq.(\ref{ooo}), $o_+(\bar\tau)$, $o_-(\bar\tau)$ and $o(\bar\tau)$ are plotted in Fig.(\ref{ooofig}). It is straightforward to find:
\begin{equation}
  \begin{split}
    & o_+ (\bar\tau) \in (0,+\infty) ,\\
    & o_- (\bar\tau) \in (\pi,o_{-u}) ,\\
    & o (\bar\tau) \in \left(o_l,\frac{\pi}{2}\right) ,\\
    & \bar\tau \in (0,\bar\tau_c) ,
  \end{split}
\end{equation}
where $o_{-u} = 4.4934094...$ and $o_l = 0.86033359...$ are the roots of $o_-\cot{o_-} = 1$ and $o\tan{o} = 1$ respectively, and $\bar\tau_c = \pi/\sqrt{2\sqrt{3}}$. It is straightforward to extract the asymptotic behavior of impact phases for the extreme values of $\bar\tau$. For $\bar\tau \to 0$, to two dominant terms, we find:
\begin{equation}
  o_z \to \nu_z^0 \bar\tau^{z+1} \left(1 + z\nu_z^1 \bar\tau^4\right), \quad
  o \to \nu_0 \left(1 + \nu_1 \bar\tau^4\right),
  \label{redfreqexp0}
\end{equation}
where
\begin{equation}
  \begin{split}
    & \nu_+^0 = \sqrt{o_l^{-2} - o_{-u}^{-2}}, \quad
    \nu_-^0 = o_{-u}, \quad
    \nu_0 = o_l, \\
    & \nu_+^1 = \frac{1}{2} \left(o_l^{-2} - o_{-u}^{-2}\right) o_l^{-2} \\
    & \qquad\qquad\quad
    - \frac{1}{3} \left(\left(o_l^2 + 2\right)^{-1} o_l^{-2} 
    - o_{-u}^{-3}\right),  \\
    & \nu_-^1 = \frac{1}{3} \left(o_l^{-2} - o_{-u}^{-2}\right) o_{-u}^{-1}, \\
    & \nu_1 = \frac{1}{3} \left(o_l^{-2} - o_{-u}^{-2}\right) 
    \left(o_l^2 + 2\right)^{-1}.
  \end{split}
  \label{nus}
\end{equation}
For $\bar\tau \to \bar\tau_c$, to the lowest non-constant term, we find:
\begin{equation}
  \begin{split}
    & o_+ \to \frac{4}{3}\epsilon^{-1}, \quad
    o_- \to \pi \left(1 + \frac{3}{4}\epsilon \right), \\
    & o \to \frac{\pi}{2} \left(1 - \frac{3}{4}\epsilon \right),
  \end{split}
  \label{redfreqexp1}
\end{equation}
where $\epsilon = -(\bar\tau-\bar\tau_c)/\bar\tau_c$.

In this section, we have omitted $\phi$ superscripts in: $w_{zz'}^{\phi s}$, $w_z^{\phi p}$, $w^{\phi d}$, $u_z^{\phi s}$, $\omega_z^{\phi s}$, $\omega^{\phi d}$, $a_-^{\phi}$, $\lambda_z^\phi$, $o_z^\phi$, $o^\phi$, $o_{-u}^{\phi}$ and $o_l^{\phi}$.

\subsection{Model parameterization of the collisionless solution}
\label{modparam}

So far, we have resolved the joining conditions in the coronal and sagittal sectors in terms of the impact phases $o_z^\phi$, $o^\phi$ and $o_-^\theta$. In the axial sector, out of two joining conditions one (the first in Eq.(\ref{axialjoinconds1})) can always be trivially satisfied, while the other (Eq.(\ref{axialjoincond2})) remains to be solved. To complete the SML solution, we need to select physically admissible values of the model parameters that reproduce the impact phases and solve the remaining joining condition, Eq.(\ref{axialjoincond2}). That can be done, in fact, for any $\bar\tau \in (0,\bar\tau_c)$, as we show in this section.

The physically realizable model parameter values must satisfy the following constraints. The parameters $l_i$, $l_h$, $l_t$, $m_i$ and $I_{i\alpha}$, where $i\in\{1,2\}$ and $\alpha \in \{\phi, \theta, \psi \}$, must all be non-negative. In addition, the moments of inertia must respect the triangle inequality: $I_{i\alpha_1} + I_{i\alpha_2} \ge I_{i\alpha_3}$, for all distinct $\alpha_i$.

In Sec. \ref{solutionconstrs} we provided a counting argument showing that at least one model parameter needs to be tuned to obtain a collisionless solution. In the linear dynamics limit the sagittal sector solution is defined up to an arbitrary scale $\varepsilon$. Setting this scale consumes an additional free parameter. Satisfying the ULI constraint takes another free parameter. Solving for an arbitrary $\bar\tau$ value consumes yet another free parameter. In total, we now need to adjust at least four model parameters.

We consider two parameterization prescriptions: a simpler (restricted) case and a more general case. We start by considering the restricted case, where we set to zero as many moments of inertia as possible, to simplify consideration: $I_{1\phi}=I_{2\phi}=I_{1\theta}=0$. The restricted case solution is then generalized to the general case of positive moments of inertia. While both cases are not totally general, they offer useful analytical insights about the extent of the domain of admissible model parameters. Refer to App. \ref{modparamapp} for more details.

\subsubsection{Restricted case parameterization}
\label{restrcase}

First, we find $l_h$, $l_1$ and $\tau$, given a solution of the $\phi$ sector equations Eq.(\ref{ooo}). For compactness, we present the results in dimensionless units by setting $g = m_2 = l_t = 1$:
\begin{equation}
  l_h = \kappa_2 \left(\kappa_1\sqrt{m_1} + 1\right), 
  \quad
  l_1 = l_h \frac{\kappa_1}{\sqrt{m_1}},
  \label{lhl1}
\end{equation}
where the universal functions $\kappa_i(\bar\tau)$ are completely independent of the model parameterization:
\begin{equation}
  \kappa_1 = \sqrt{-\bar\lambda_+^\phi \bar\lambda_-^\phi}, 
  \quad
  \kappa_2 = \left(\bar{a}_{+-}^\phi\right)^{-2} ,
  \label{kappa12}
\end{equation}
where $\bar\lambda_z^\phi = \lambda_z^\phi/\gamma = z(o_z^\phi(\bar\tau)/\bar\tau)^{-2}$, and $\bar{a}_{zz'}^\phi = a_{zz'}^\phi/\gamma = \Lambda_{zz'}(\bar\lambda_+^\phi, \bar\lambda_-^\phi, 1)$. For any $\bar\tau \in (0,\bar\tau_c)$ the functions $\kappa_1$ and $\kappa_2$ are positive, and so are $l_h$ and $l_1$. The condition $l_2 > 0$ is satisfied as long as $m_1 > m_c$, where
\begin{equation}
  m_c = \left\{
  \begin{split}
    0, \quad \kappa_2 \ge 1, \\
    \left(\frac{1-\kappa_2}{\kappa_1\kappa_2} \right)^2, \quad \kappa_2 < 1,
  \end{split}
  \right.
  \label{mc}
\end{equation}
The impact time is
\begin{equation}
  \tau = \bar\tau \kappa_2^{\frac{1}{4}}
  \label{tau}
  .
\end{equation}
 Next, we need to find $I_{2\theta}$. Notice, its value does not affect $\phi$ sector. The equation Eq.(\ref{osagittal}) (which is a consequence of the ULI constraint) can be solved for a positive $I_{2\theta}$ given any $\bar\tau \in (0,\bar\tau_c)$. Indeed, if $I_{2\theta}=0$ then $o_-^\theta = o_-^\phi > \pi$, and if $I_{2\theta}\to +\infty$ then $o_-^\theta \to 0$. Therefore, there exists $I_{2\theta} > 0$ that solves $o_-^\theta = \pi$:
\begin{equation}
  I_{2\theta} = 
  \frac{\tau^2}{\pi^2} + 
  \left(\frac{\tau^2}{\pi^2} \kappa_2^{-1} + \kappa_1^2 + 1 \right)^{-1} - 1
  \label{i2thetarestr}
  .
\end{equation}
To respect the triangle inequalities on the moments of inertia, one should set $I_{1\psi} = 0$ and $I_{2\psi} = I_{2\theta}$. Coincidentally, this choice of $I_2$ parameters also satisfies the axial plane joining condition Eq.(\ref{axialjoincond2}), as $b_1 = 0$ now. Thus, the restricted case SML solution is parameterized by 4 independent parameters: 2 model parameters ($d$ and $m_1$), and 2 non-model parameters ($\bar\tau$ and $\varepsilon$). 

To restore the presented expressions to dimensionful notations one needs to replace all mass $m_a$, length $l_a$, moment of inertia $I_a$ and time $\tau$ notations with $m_a/m_2$, $l_a/l_t$, $I_a/m_2l_t^2$ and $\tau\sqrt{g/l_t}$ respectively, where $a$ stands for any subscript.

\subsubsection{General case parameterization}
\label{gencase}

To consider the general case of positive moments of inertia, we parameterize them as $I_{1\alpha} = c_1^\alpha \tilde\mu_2$ and $I_{2\alpha} = c_2^\alpha m_2l_t^2$. While $I_{1\alpha}$ is now parameterized implicitly (since $\tilde\mu_2$ is not fixed), this type of analysis still gives us a good idea about the range of $I_{1\alpha}$ values, for which an admissible parameterization of the SML solution exists. Let us define
\begin{equation}
  \begin{split}
    & \kappa'_1= \sqrt{\frac{\kappa_1^2+1}
    {\left(1+c_1^\phi\right)\left(1+c_2^\phi\right)} - 1} ,\\
    & \kappa'_2 = \kappa_2\left(1+c_2^\phi\right)^{2} .
  \end{split}
  \label{newkappa}
\end{equation}
If we now replace $\kappa_i$ with $\kappa'_i$ in Eqs.(\ref{lhl1},\ref{mc},\ref{tau}), we obtain the general case solution for $l_h$, $l_1$ $m_c$ and $\tau$, (see App. \ref{modparamapp} for more details). The solution is defined for those values of $\bar\tau$ and $c_i^\phi$ that do not turn the expression under the square root in Eq.(\ref{newkappa}) negative.
The general case formula for $I_{2\theta}$ involves more than just replacing $\kappa_i$ with $\kappa'_i$:
\begin{equation}
  I_{2\theta} = 
  \frac{\tau^2}{\pi^2} + 
  \left(\frac{\tau^2}{\pi^2} {\kappa'_2}^{-1} 
  + \left({\kappa'_1}^2 + 1\right)\left(1+c_1^\theta\right) \right)^{-1} - 1
  \label{i2thetagen}
  .
\end{equation}

Let us investigate the case of $I_{2\phi} = I_{2\theta} = \zeta$. In this case $\zeta$-dependence drops out of Eq.(\ref{i2thetagen}) completely, and we can rewrite it as:
\begin{equation}
  c_1^\theta = \left(1+c_1^\phi\right)
  \frac{(1-\xi^2)^{-1}-\xi^2\kappa_2^{-1}}{\kappa_1^2+1}-1,
  \label{c1theta}
\end{equation}
where $\xi = \bar\tau \kappa_2^{1/4}/\pi$. The equation has the form $c_1^\theta = (1+c_1^\phi)\Phi(\xi,\kappa_1,\kappa_2)-1$; it is easy to verify (see App. \ref{modparamapp} for details) that $\Phi(\xi,\kappa_1,\kappa_2) > 1$ as long as $\xi < 1$, and therefore, $c_1^\theta > c_1^\phi$. Thus, we have an explicit prescription, in terms of $c_i^\alpha$, for the model parameterization in the general case for any $\bar\tau$, as long as $(1+c_1^\phi)(1 + \zeta) < \kappa_1^2 + 1$.

\begin{figure}
  \includegraphics[width=1\columnwidth]{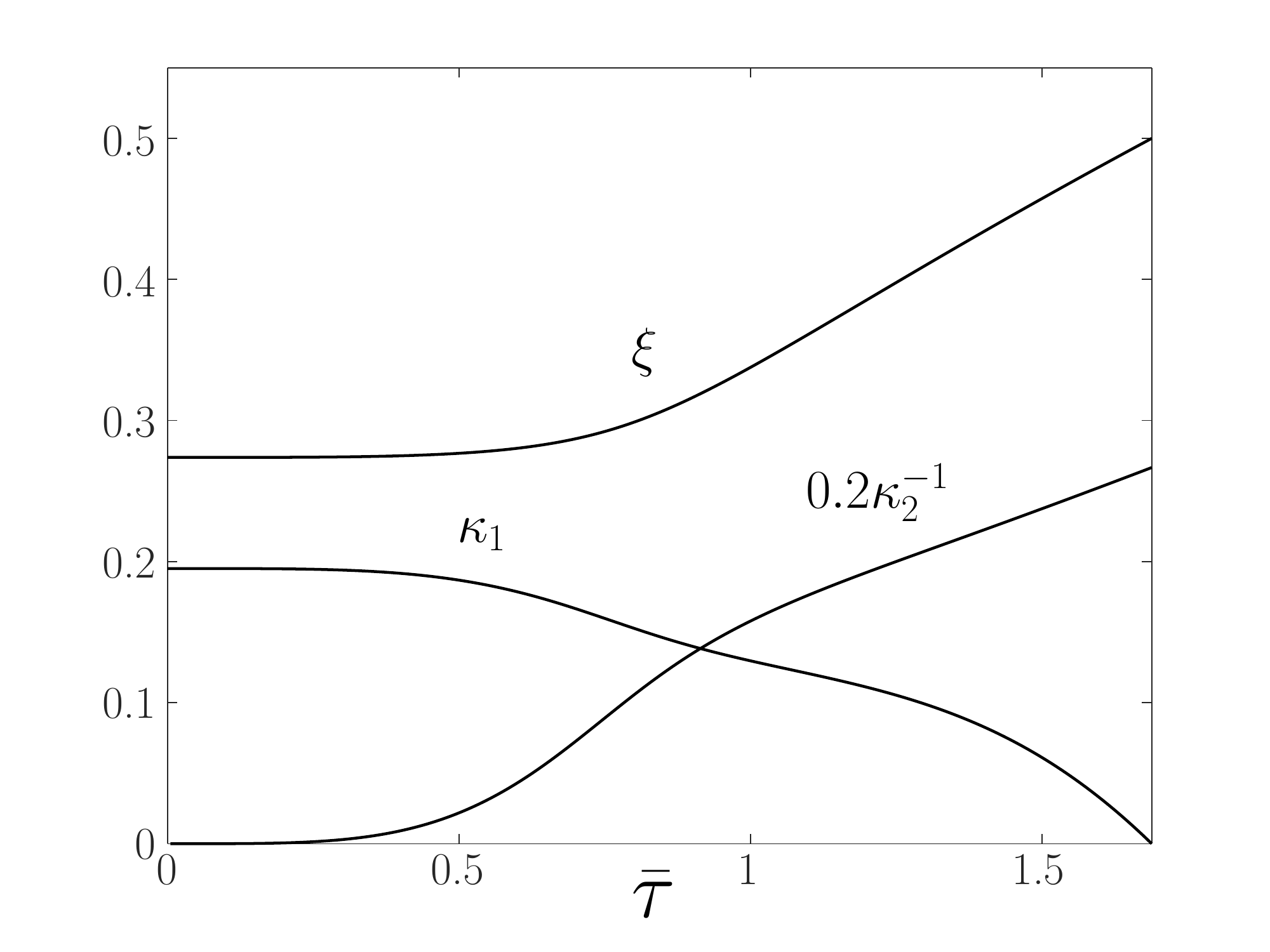}
  \caption{Universal functions $\kappa_i(\bar\tau)$ and $\xi(\bar\tau)$ (plotted for $\bar\tau\in(0,\bar\tau_c)$) are independent of the model parameterization. Because $\kappa_2$ is unbounded from above, it is plotted as $0.2/\kappa_2$.
}
  \label{kappaxifig}
\end{figure}

To have a more quantitative assessment of the admissible model parameter domain, we present the asymptotic expansion of $\kappa_i$ to the lowest non-constant term, as well as the expansion of $\bar\lambda_z$ to the lowest order needed to establish the expansion of $\kappa_i$.   
For $\bar\tau \to 0$, we find from Eqs.(\ref{redfreqexp0},\ref{nus},\ref{kappa12}):
\begin{equation}
  \bar\lambda_z \to z{\nu_z^0}^{-2} \bar\tau^{-2z} \left(1 - 2z \nu_z^1 \bar\tau^4\right)
\end{equation}
and
\begin{equation}
  \begin{split}
    & \kappa_1 \to \left(\nu_+^0 \nu_-^0 \right)^{-1}
    \left( 1 - \left(\nu_+^1 - \nu_-^1 \right)\bar\tau^4 \right), \\
    & \kappa_2 \to \nu_0^4 \bar\tau^{-4}.
  \end{split}
\end{equation}
For $\bar\tau \to \bar\tau_c$, we find from Eqs.(\ref{redfreqexp1},\ref{kappa12}):
\begin{equation}
  \bar\lambda_+  \to \frac{3\sqrt{3}}{32} \pi^2 \epsilon^2, \quad
  \bar\lambda_- \to -\frac{1}{2\sqrt{3}} \left(1 - \frac{7}{2}\epsilon \right)
\end{equation}
and
\begin{equation}
  \kappa_1 \to \frac{\sqrt{3}}{8} \pi\epsilon, \quad
  \kappa_2 \to \frac{3}{4} \left(1 + \epsilon \right).
\end{equation}
The functions $\kappa_i$ and $\xi$ are plotted in Fig(\ref{kappaxifig}). We have verified that $\kappa_i$ are monotonically decreasing functions, while $\xi$ is a monotonically increasing function of $\bar\tau$. The function $\xi$ changes from $\xi \to \nu_0/\pi$ at $\bar\tau \to 0$ to $\xi \to 1/2$ at $\bar\tau \to \bar\tau_c$. Thus the required condition $\xi < 1$ is satisfied for all $\bar\tau$. Notice that $\tau \to \nu_0\sqrt{1+c_2^\phi}$ remains finite for $\bar\tau\to 0$.

The axial joining condition Eq.(\ref{axialjoincond2}) is equivalent to $b_1^\T u_-^\theta = 0$, from where we find, for $I_{2\phi} = I_{2\theta} = \zeta$: $I_{2\psi} = \zeta\left(1-u_{-(1)}^\theta/u_{-(2)}^\theta\right)$. Since $0 < u_{-(1)}^\theta < u_{-(2)}^\theta$, (see App. \ref{modparamapp} for details), we have $0 < I_{2\psi} < \zeta$, which satisfies the triangle inequality. To satisfy it for $I_1$, we choose $I_{1\psi}$ in the range $|I_{1\phi} - I_{1\theta}| \le I_{1\psi} \le I_{1\phi} + I_{1\theta}$. Thus, the general case SML solution is parameterized by 7 independent parameters: 5 model parameters ($d$, $m_1$, $c_1^\phi$, $c_2^\phi$ and $I_{1\psi}$), and 2 non-model parameters ($\bar\tau$ and $\varepsilon$). 

We have demonstrated that both the restricted and general case of the model parameterization satisfies all the joining equations and the ULI constraint, and therefore realizes a collisionless solution in the SML.

It is instructional to evaluate the effective number of the tuned model parameters in the restricted and general cases. We ignore the nonholonomic constraints, even when they become degenerate, as in the restricted case. We also ignore the general case constraint $c_2^\phi = c_2^\theta$, which can be replaced by a nonholonomic constraint. Then, in both the restricted and general case we only need to tune two additional model parameters to satisfy the ULI constraint and the axial joining condition. It makes four parameters in total (together with $l_h$ and $l_1$), as expected. 

\subsection{Spring supported torso solution}
\label{duality}

\newcommand{\ke}{\mathrm k}
\newcommand{\y}{\mathrm y}
\newcommand{\x}{\mathrm x}

In this section we consider a version of the original model endowed with torsion springs at the hip joint. This modification enables consideration of a collisionless gait with the torso located above the hip joint, which can be regarded as a more anthropomorphic-looking gait.

The hip springs are modeled by adding the elastic energy term 
\begin{equation}
V_e = \frac{\ke}{2} q_t^\T q_t
\end{equation}
to the potential energy $V$, where $\ke$ is the spring constant. This changes the potential term $G$ by $\nabla V_e$ in the equations of motion Eq.(\ref{eqmotion}), without affecting $H$ and $C$. In the SML, only $G_1$ is affected and only in $\alpha\in\{\phi,\theta\}$ sectors, cf. Eqs(\ref{lindyneqmot},\ref{hg0g1alpha}):
\begin{equation}
  G_{1\alpha} = g 
  \begin{bmatrix}
    -\mu_1 &  m_2 l_t \\
    m_2 l_t & m_2 l_t + \frac{\ke}{g}
  \end{bmatrix}
  .
\end{equation}
Note, the same equations can be used for describing both the hanging and standing torso arrangements, by choosing $l_t < 0$ for the standing torso case. As long as the choice of model parameters does not affect the signs of the eigenvalues of $M$ in Eq.(\ref{Mmat}), all the analysis in terms of the impact phases in Sec. \ref{sagplanesol} and \ref{corplanesol}, as well as the universal functions, remain unchanged. Only the model parameterization step needs to be reconsidered. Rather than reproducing a somewhat tedious analysis of Sec. \ref{modparam} for an arbitrary $\ke$ of the spring-endowed model, we take a different and simpler route. We establish an up-down torso duality, that connects solutions at different $\ke$ values and $l_t$ signs. We then use this duality to resolve the parameterization of the spring-supported standing torso solution in terms of the already solved springless hanging-torso solution parameterization.

We switch to dimensionless units $g = m_2 = 1$, $l_t = z = \pm 1$ with $z = -1$ corresponding to the standing torso setup. We generalize $\tilde S$ so that it covers both values of $z$ and matches $z = 1$, $\ke = 0$ case of Sec. \ref{sml} 
\begin{equation}
  \tilde S = 
  \begin{bmatrix}
    1 &  0 \\
    \y & z
  \end{bmatrix}
  ,
  \label{tS}
\end{equation}
where $\y = 1/(z+\ke)$ is chosen to make $\tilde G_{1\alpha}$ diagonal. We only use $\alpha$ for $\alpha\in\{\phi,\theta\}$ sectors. We use the prime to denote the quantities in the dual setup, corresponding to $z' = -z$. The duality relation is induced by requiring the parameters of the dual setup to be selected so that $\tilde H_\alpha$ and $\tilde G_{1\alpha}$ are independent of $z$. Importantly, (one can easily verify that) the joining conditions are independent of $z$ in the $\alpha$ sectors. Therefore, provided the dual parameter values are physically admissible, the solution $\tilde q_\alpha' = \tilde q_\alpha$ is a valid collisionless solution, as it satisfies both the equations of motion and the joining conditions of the dual setup. In the $\psi$ sector, the joining conditions do depend on $z$, requiring a proper selection of $I_{2\psi}$, as we explain at the end of this section.

We now present the explicit form of the duality transformation for all the model parameters, except $I_{2\psi}$. For simplicity, we keep $I_{i\alpha}$ fixed, only allowing $\ke$, $m_1$, $l_1$ and $l_h$ to vary with $z$. Under these restrictions we find for the duality transformation (see App. \ref{dualityapp} for details)
\begin{equation}
  \begin{split}
    & z' = -z , \\
    & \ke' = \ke + 2z , \\
    & m_1' = \frac{\delta_1^2}{\delta_2} , \quad
    l_1' = \frac{\delta_2}{\delta_1} , \\
    & l_h' = l_h + 2z(1+I_{2\alpha}) ,
  \end{split}
  \label{dualtransf}
\end{equation}
where we have defined
\begin{equation}
  \begin{split}
    & \delta_1 = m_1l_1 - 2z(2+I_{2\alpha}), \\
    & \delta_2 = m_1l_1^2 - 4z(2+I_{2\alpha})(l_h + z(1+I_{2\alpha})).
  \end{split}
\end{equation}
For $m_1'$ and $l_1'$ to be physically admissible, the positivity of $\delta_i$ must be ensured, leading to the condition on $m_1$
\begin{equation}
  m_1 >  \left(\max{\{\sqrt{m_{c}}, \sqrt{m_{c1}}, \sqrt{m_{c2}}\}}\right)^2,
  \label{dualmccond}
\end{equation}
where
\begin{equation}
  \begin{split}
    & \sqrt{m_{c1}} = \left(\sqrt{1+8\frac{2+I_{2\alpha}}{\kappa_2}}-1\right)
    \frac{1+z}{4\kappa_1} , \\
    & \sqrt{m_{c2}} = 2\frac{2+I_{2\alpha}}{\kappa_1^3\kappa_2}\left(\sqrt{1+\frac{1+I_{2\alpha}}{2+I_{2\alpha}}\kappa_1^2}+z\right)-\frac{1}{\kappa_1} .
  \end{split}
  \label{mc12}
\end{equation}

Note that $\mu_1$, $\mu_2$ and $\y$ do not change under the duality transformation.

Let us now consider the selection of $I_{2\psi}$, (which has no affect on the $\alpha$ sectors). In the restricted case, Eq.(\ref{axialjoincond2}) is trivially satisfied (see Sec. \ref{restrcase}), so $I_{2\psi}' = I_{2\psi}$. In the general case, reproducing the analysis of Sec. \ref{gencase} for general $\y$ and $z$, we find:
\begin{equation}
  I_{2\psi} = \zeta \frac{1-\x\y}{1+\x(z-\y)},
  \label{duali2psi}
\end{equation}
where $\x = u_{-(1)}^\theta / u_{-(2)}^\theta$, $0 < \x < \beta^{-1} < 1$ (see App. \ref{modparamapp}) and $0 < \y \le 1$. To satisfy the triangle inequality on $I_{2\psi}$ we need $0 < I_{2\psi} < 2\zeta $, which is satisfied for any $\x$ if $z=1$ and for $\x < 1/(2+\y)$ if $z=-1$. From there a condition on $m_1$ follows (see App. \ref{dualityapp}):
\begin{equation}
  \sqrt{m_1} > \sqrt{m_{c3}} = 
  \left(\frac{2+\y}{\sqrt{\kappa_2}}-1\right) \frac{1-z}{2\kappa_1} .
  \label{mc3}
\end{equation}

To obtain a standing torso solution parameterization in the general case, one can start with any valid springless model parameterization (as presented in Sec. \ref{modparam}), provided that the conditions Eq.(\ref{dualmccond},\ref{mc3}) are met. One then uses the duality transformation Eq.(\ref{dualtransf}) (with $z = 1$ and $\ke = 0$) and Eq.(\ref{duali2psi}) (with $z = -1$ and $\y=1$) to compute all the model parameters in the dual setup. The restricted case is even simpler: only Eq.(\ref{dualmccond}) needs to be enforced and instead of using Eq.(\ref{duali2psi}) one sets $I_{2\psi}' = I_{2\psi}$.

Importantly, satisfying the conditions on $m_1$ poses no practical difficulty. It is straightforward to verify, that for a given springless solution, the conditions Eq.(\ref{dualmccond},\ref{mc3}) can always be satisfied by keeping $\bar\tau$ and $I_{i\phi}$ fixed, while increasing $m_1$ and computing all other parameters as prescribed by the springless model parameterization procedure. Note, we do not need to know $\bar\tau'$, as the duality approach bypasses the use of the universal functions in the dual setup.

\section{General solution}
\label{gensol}

Let $\chi$ include all the adjustable initial conditions on the generalized coordinates and the impact times, and let $\eta$ include all the adjustable model parameters:
\begin{equation}
  \begin{split}
    & \chi = [\phi^s_l,\phi^s_t,\dot\theta^s_l,\dot\psi^s_l,\dot\theta^s_t,\dot\theta^d_l,\dot\phi^d_t,\dot\theta^d_t,t_s,t_d]^\T, \\
    & \eta = [l_h,l_1,d,m_1,\eta_I^\T]^\T,
  \end{split}
\end{equation}
where $\eta_I$ is the triangle-inequality-respecting parametrization of the moments of inertia, see App. \ref{constrgraddesc} for details. Thus, $x \equiv [\chi;\eta]$ includes all the parameters that can be adjusted while searching for a collisionless gait solution. Let $h(x)$ be 11-dimensional vector combining the joining conditions from Eqs.(\ref{impactconds},\ref{matchconds}). We are interested in finding solutions of $h(x) = 0$, possibly with certain desired properties of $x$. A general solution can only be found numerically by an iterative procedure, such as Newton's method. 
In a complex high-dimensional problem without a good initial guess on $x$, this iterative procedure is not even guaranteed to converge to $h = 0$, let alone in a particular region with desired properties. Therefore, it is highly preferable that the initial guess $x_0$ be already close to the solution, that is $h(x_0) \approx 0$. Our SML solution provides such $x_0$. We then evolve it to a desired region in the parameter space using a constrained gradient descent technique. It involves optimizing an objective function $f(x)$ of our choice, while staying close to a valid solution $h(x_i) \approx 0$ at any step $i$ in the course of optimization. The constrained gradient descent update rule is (see App. \ref{constrgraddesc} for details):
\begin{multline}
  x_{i+1} = x_i -\lambda_x^{-1}\left(I_x - {h'(x_i)}^+h'(x_i) \right)
  \nabla{f}(x_i) \\ - {h'(x_i)}^+h(x_i)
  \label{cgd}
\end{multline}
where the shorthand notation $h'$ stands for the matrix $h'_{(ij)}\equiv\partial{h}_{(i)}/\partial{x}_{(j)}$, and $h'^+$ is a pseudo-inverse of $h'$, (defined for a matrix $m$ as $m^+ = m^\T \left(mm^\T\right)^{-1}$). The second term in the right hand side of the above equation is the projection of the gradient $\nabla f$ onto the tangent space of $h = 0$, while the last term enforces $h = 0$. The coefficient $\lambda_x$ controls the gradient descent rate and $I_x \equiv I_{\dim(x)\times\dim(x)}$.

We use the general case parameterization for $x_0$, as prescribed in Sec. \ref{gencase}, with $d$ set to a small value. We then evolve $x$ away from the SML using Eq.(\ref{cgd}) with $f$ encoding certain soft constraints, for example, favoring larger values of $d$ and penalizing excessively small values of $l_h$, $l_1$ and $I_\phi$. We also penalize small values of $|\dot\psi_l(0)|$ to prevent collapse onto a purely coronal movement solution (side-to-side rocking with zero step length, i.e. $\psi_l = \theta_l = \theta_t = 0$). The optimization procedure was implemented using a freely available GNU Octave software \cite{octave}, and we found the optimization to be fairly straightforward.

Let the integers $(k_\phi,k_\theta)$ be the numbers of oscillations in the coronal and sagittal sectors per one collisionless walking cycle. (Due to the imposed symmetries, the number of oscillations in the $\phi$ and $\theta$ sectors between two nearby symmetry points of the same type must be half-integer and integer respectively, therefore $k_\phi$ and $k_\theta$ must be respectively odd and even.) We call such solution a $(k_\phi,k_\theta)$-mode. So far in Sec. \ref{smlsolution} we have analyzed the (3,4)-mode. As one stays near a valid solution during optimization, the discrete numbers $(k_\phi,k_\theta)$ do not change, (perhaps with exception of certain points in the space of collisionless trajectories). To explore different general modes, one should seed the optimization with different SML modes. In the SML, it should be possible to analyze all the modes analytically, similarly to the (3,4)-mode. This goes beyond the scopes of the paper though, as the modes with high number of oscillations seem less interesting for the purpose of walking. In the general regime we focus exclusively on the (3,4)-mode, which is the simplest SML mode compatible with the ULI constraint.

We have evolved the (3,4)-mode from an initial general case SML solution (see Sec. \ref{gencase}) toward a non-small feet separation solution, following the outlined optimization procedure. We then truncated all but one of the model parameters to two significant digits and ran a few more iterations of the optimization step Eq.(\ref{cgd}) with $\eta = [l_h]$ and $\nabla f = 0$, thus only optimizing over $l_h$ to enforce $h = 0$. Our claim that a single model parameter suffices to tune a general collisionless mode is evidenced by a prompt drop of $\lVert h \rVert$ to the numerical zero. All the results presented below are for this particular solution, encoded by the following model parameter values: $l_h = 1.0941669$, $l_1 = 0.16$, $d = 0.15$, $m_1 = 0.19$, $I_{1\phi} = 0.00002$, $I_{1\theta} = I_{1\psi} = 0.00032$, $I_{2\phi} = 0.0000092$, $I_{2\theta} = I_{2\psi} = 0.019$, (we use $g = m_2 = l_t = 1$). The corresponding values of $\chi$ are given in App. \ref{miscrels}.

\begin{figure}
  \includegraphics[width=1\columnwidth]{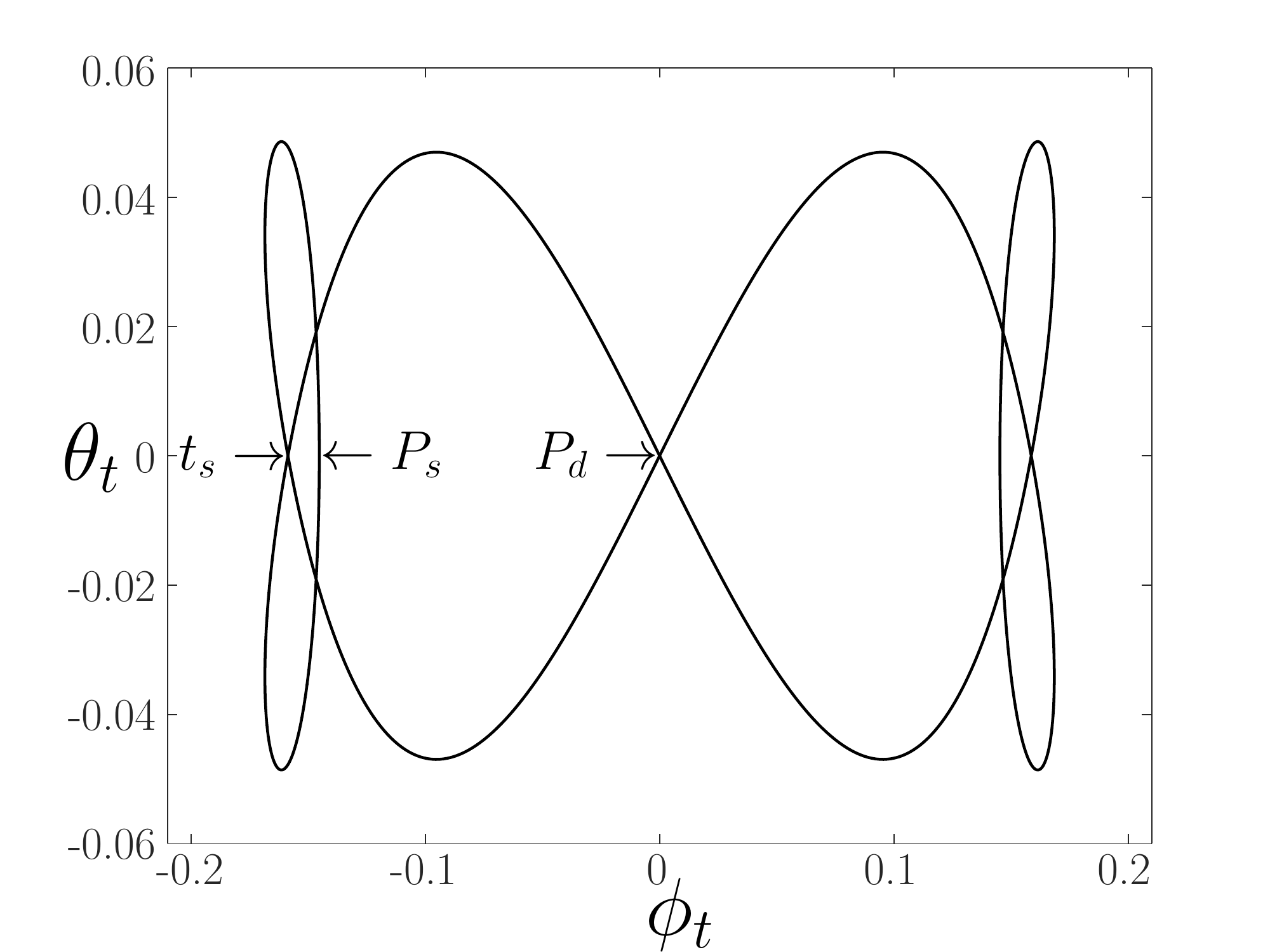}
  \caption{The parametric plot of $\theta_t(t)$ vs $\phi_t(t)$ shows the torso's trajectory relative to the legs over one walking cycle for the (3,4)-mode. The Lissajous-like curve features 3 coronal and 4 sagittal oscillations. The symmetry points $P_s$ and $P_d$, and the impact point (indicated here as $t_s$) are denoted in the left half of the plot.
}
  \label{torsotrajfig}
\end{figure}

The torso's trajectory relative to the legs for the (3,4)-mode is plotted in Fig.(\ref{torsotrajfig}). The trajectory is loosely reminiscent of a Lissajous curve, as it is a closed contour of two superimposed oscillating motions, but only the sagittal oscillations are simple harmonic (and only in the SML).

\begin{figure}
  \includegraphics[width=1\columnwidth]{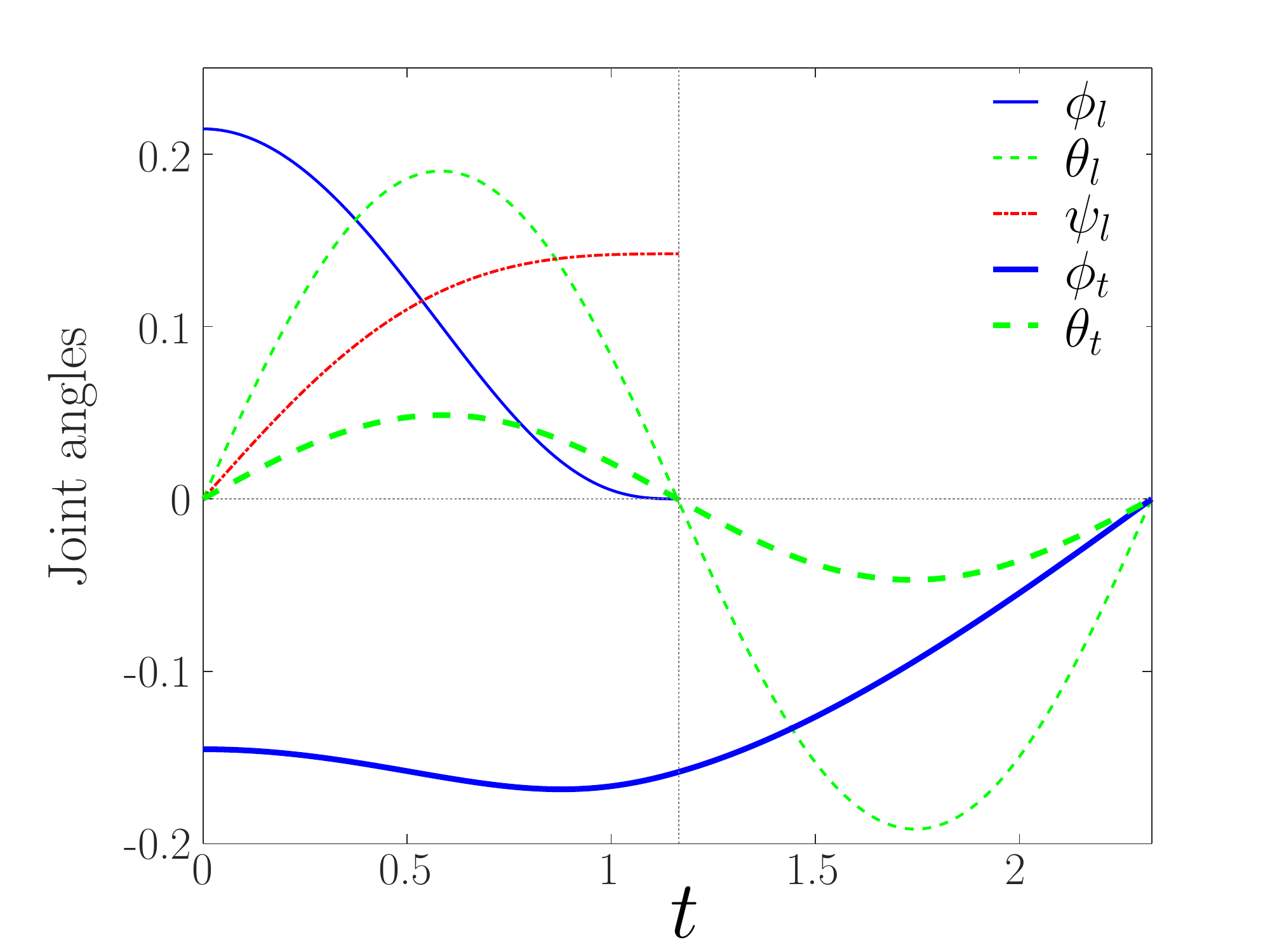}
  \caption{(Color online) Collisionless trajectory components are plotted between $P_s$ and $P_d$ with the following nomenclature: solid (blue), dashed (green) and dash-dotted (red) lines represent the $\phi$, $\theta$ and $\psi$ sectors respectively, while $q_l$ and $q_t$ are indicated by thin and bold lines respectively. Vertical line is the impact moment separating phases I and II. The components $\theta_l$, $\psi_l$ and $\theta_t$ vanish at $P_s$ by symmetry. The components $\theta_l$, $\phi_t$ and $\theta_t$ vanish at $P_d$ by symmetry. Only $\phi_l$ vanishes at the impact. }
  \label{qpspdfig}
\end{figure}

\begin{figure}
  \includegraphics[width=1\columnwidth]{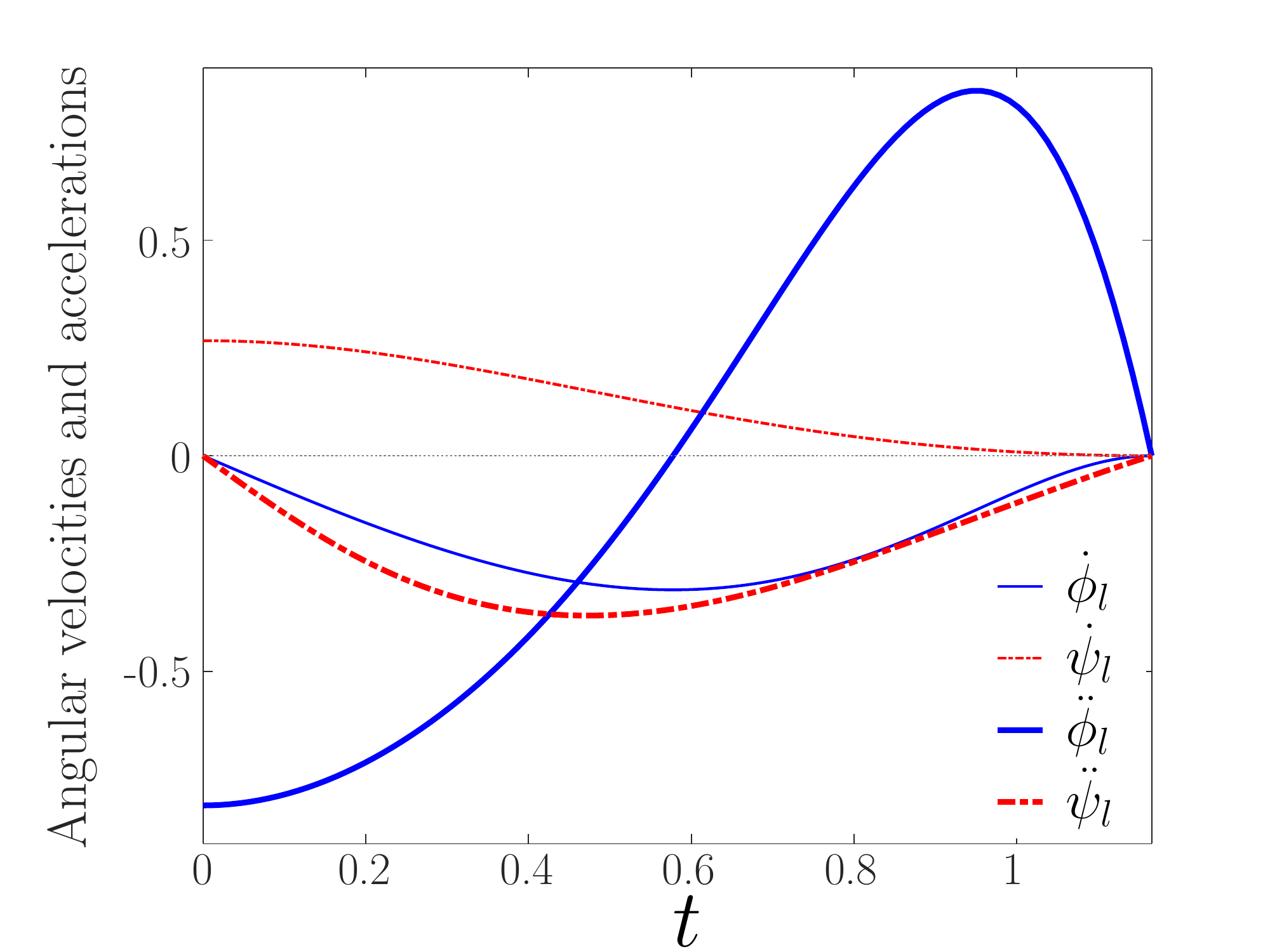}
  \caption{(Color online) The first and second time derivatives (angular velocities and accelerations) that need to vanish at the impact for the solution to realize a collisionless conventional walking: $\dot\phi_l(t)$ (solid blue), $\dot\psi_l(t)$ (dash-dotted red), $\ddot\phi_l(t)$ (bold solid blue) and $\ddot\psi_l(t)$ (bold dash-dotted red) are plotted between $P_s$ and impact. All the shown derivatives vanish at the impact.}
  \label{dqddqfig}
\end{figure}

A collisionless trajectory is fully specified by its definition between nearby symmetry points. In Fig.(\ref{qpspdfig}), we plot $q(t)$ between $P_s$ and $P_d$, which constitutes a quarter of the cycle. A complete walking cycle trajectory can be obtained by unfolding the plot around the symmetry points with appropriate symmetry transformations. Note, only $\phi_l$ vanishes at the impact, $\theta_l$ and $\theta_t$ remain finite, as the ULI constraint is not enforced in the general solution. Two components, $\phi_l$ and $\psi_l$, become inactive in a two-dimensional impact. Those components must respect the impact conditions of vanishing first and second time derivatives. This is explicitly demonstrated in Fig.(\ref{dqddqfig}) by plotting the derivatives for $t \in (0,t_s)$. While the velocities and accelerations of $\phi_l$ and $\psi_l$ vanish at the impact, their jerks remain finite \cite{chatterjee2002persistent}.

\begin{figure}
  \includegraphics[width=1\columnwidth]{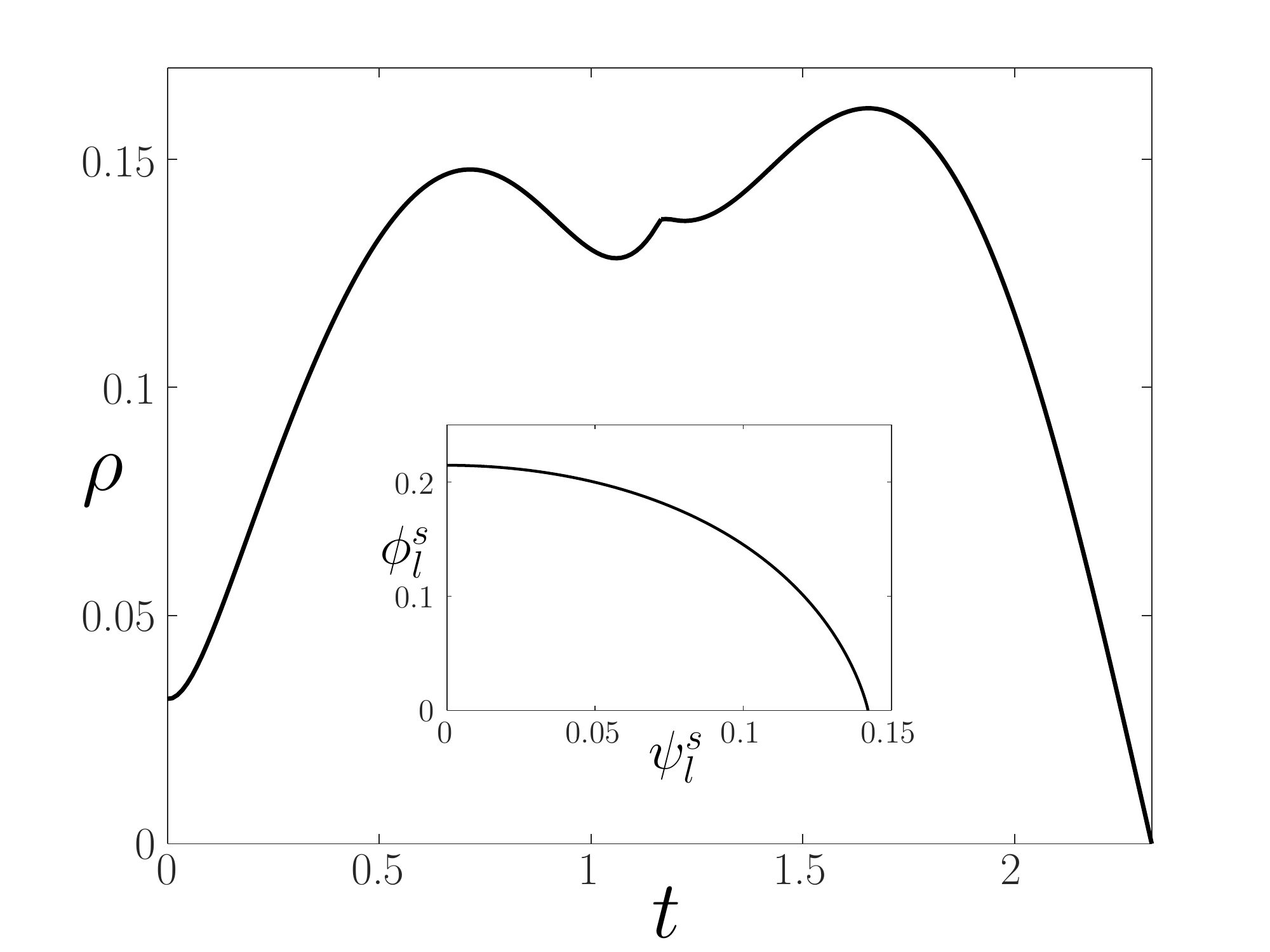}
  \caption{ Main panel: the minimum friction coefficient $\rho(t)$, sufficient to prevent sliding, is plotted between $P_s$ and $P_d$. Inset: the parametric plot of $\phi_l(t)$ vs $\psi_l(t)$ for $t\in(0,t_s)$ shows the swing foot trajectory. The finite slope at the impact indicates finite $\rho(t_s)$.
}
  \label{fcfig}
\end{figure}

In the inset of Fig.(\ref{fcfig}) we plotted the swing foot trajectory, to illustrate that no ground penetration or scuffing occurs in the swing phase. Let $\rho(t)$ be the minimal friction coefficient sufficient to prevent foot sliding at time $t$, (see also App. \ref{miscrels}). The finite angle of the trajectory at the impact (equal to ${\rm atan2}(\dddot\phi_l(t_s),\dddot\psi_l(t_s))$) implies a finite angle of the foot-strike with the ground, indicating finite $\rho(t_s)$. The main panel of Fig(\ref{fcfig}) shows $\rho(t)$ reaching maximum value of about $0.161$ in phase II. Note, the forces are continuous across the impact due to the enforcement of the impact conditions, while the jerks are not, as manifested by a kink in $\rho(t)$ at $t_s$. We have verified that the ground reaction force remains non-negative throughout the cycle, thus confirming that the computed collisionless gait represents conventional walking.

The gait features the step length of about $0.085$, which is 7.7\% of the model height or 27.4\% of the feet separation. This may seem not a lot for a normal walking, but given overall simplicity of the model and fixed feet separation, it does not look too little either. We can liken it to a penguin walk, whose feet's range is also somewhat restricted (in comparison to many bipedal animals), and whose wobbling walk also involves significant torso movements.

\section{Related work and discussion}
\label{discussion}

The optimization of robot's design and control policies for energy efficiency is ubiquitous in the field of legged locomotion. Yet, relatively few studies have addressed the possibility of complete elimination of the ground collision losses in passive walkers, even though it offers obvious benefits of greatly (indefinitely, barring the joint and air friction) extending the walking range of a robot. While the current record-holding robot \cite{bhounsule2014low} has narrow specialization, there is no fundamental reason for why an extreme walking efficiency and practical versatility should be mutually exclusive in robots.

The impact conditions that need to be satisfied by a collisionless conventional walking trajectory were formulated in Refs. \cite{reddy2001passive,chatterjee2002persistent}. The actual models chosen for the collisionless motion demonstration had relied in their design on springs. We do not view it as a critical feature, as oscillating motion can be realized by hanging pendulum-like parts as well. Springs do offer more freedom in the selection of the model's geometry, however. For example, they can be added to prop up the torso above the hip, perhaps to give the robot a more presentable look. Indeed, in Sec. \ref{duality} we extended our design to include hip springs and showed that this modification only affects the parameterization of the model, but not the universal aspects, such as the shape of $\kappa_i$. Ultimately, we believe the springs are only strictly necessary for propping up a single-link torso, as they enable an oscillatory mode of the standing torso, which is needed for the collision cancellation. Most likely, the same effect can be achieved by attaching the second link (e.g. a hanging arm) at the top of the standing torso.

A planar three-link biped model with a spring-mounted torso, potentially capable of collisionless walking was proposed in Ref. \cite{chatterjee2000small} . It was hypothesized that, with a torso, the model had sufficient number of internal degrees of freedom to eliminate collisional losses. A version of this model, with each leg connected to the torso by a torsional spring, was solved numerically in Ref. \cite{gomes2011walking}, (also in Ref. \cite{chyou2011upper}, with an arm added to the torso). However, that solution does not enforce the acceleration vanishing at foot impacts, and therefore does not represent conventional walking. Without an additional mechanism, the walker would not be able to follow the calculated trajectory, as it would be losing its contact with the ground upon a foot strike.

The symmetry points $P_s$ and $P_d$ can be viewed as a generalization of the planar symmetry points imposed in Ref. \cite{gomes2011walking}. There, an additional constraint of instantaneous support transfer was introduced, corresponding to $t_d\to 0$ (the impact merging with $P_d$) in our terminology. It turns out, this constraint is incompatible with the existence of a collisionless (conventional) walking solution \footnote{Private communication with the authors.}. The double support phase appears to be necessary for realizing collisionless walking, as was originally suggested in Ref. \cite{chatterjee2000small}.

In Sec. \ref{solutionconstrs} we presented a general counting argument for the number of model parameters $n$ that need to be tuned to realize a collisionless motion: $n$ is one less than the impact dimension. It explains in a unified manner why $n = 0$ for the hopper \cite{chatterjee2002persistent} and the (extended) rimless wheel \cite{gomes2005collisionless}, while $n = 1$ for the three-link walker and our model. Note, $n$ does not depend directly on the number of model degrees of freedom or spatial dimensions. ``Tuning'' refers to a small adjustment in the vicinity of a valid solution. It does not mean that 9 remaining (dimensionless) model parameters can assume arbitrary values. Still, in the SML we were able to explicitly parameterize a wide range of admissible model parameters by means of a set of universal functions of reduced impact time (an independent non-model parameter) and 5 independent model parameters with clear restrictions on them. We find it somewhat remarkable and credit the imposition of the ULI constraint for that.

Our main contribution in this paper is the analytical analysis of a collisionless walking solution of a three-dimensional bipedal model in the small movement limit, aided by the upright legs constraint at the impact. Similar amplitude smallness considerations have been applied to passive walking analysis on small or vanishing slopes \cite{garcia1998simplest,chatterjee2000small,garcia2000efficiency}, but not in the context of complete elimination of collisions, as far as we know. We consider it intuitively plausible that the SML collisionless spectrum is representative of the general collisionless modes in sufficiently large vicinity of the SML. In the same vein as the linear normal modes are representative of the nonlinear normal modes in conservative systems, under certain conditions \cite{lyapunov1992general}. However, we do not attempt to rigorously defend this idea. Instead, we empirically demonstrate the validity of such reasoning, by numerically extending the (3,4)-mode to non-small feet separation. 

Our analysis implies that in general $k_\phi \ge 3$, and that the (3,4)-mode is the simplest mode under the ULI constraint. Whether an even simpler mode -- (3,2)-mode -- exists in the absence of the constraint, is an open question. We think the SML-like analysis of collisionless solution looks promising for other models, potentially more complex and more anthropomorphic models. It would be interesting to apply it to the planar three-link model. Particularly, it would be interesting to see if the problem separates into a universal part (like our equations on the impact phases) and a parameterization part, or if one needs to introduce additional constraints to make the problem more tractable. It would also be interesting to consider a three-dimensional three-link model with a pelvis and spring supported torso, similar to the model introduced in Ref. \cite{faraji20173lp}. To establish a walking SML solution, it would be sufficient to consider linear dynamics in the decoupled sagittal and coronal planes, (with the sagittal plane dynamics identical to that of the planar three-link model). We conjecture that this model possesses collisonless walking modes, requiring two model parameter tuning due to its three-dimensional impacts. These are just some of the possible research directions worth pursuing.

There are many ways to reduce collisional energy losses in legged locomotion, for example, via use of light-weight legs or tendon-like springs \cite{srinivasan2011fifteen}. Yet, only the elimination of collisions themselves can completely eliminate collisional losses. This work extends our theoretical understanding of the related ideas and could improve the design of efficient walking robots.

\section{Acknowledgments}

We thank Andy Ruina and Mario Gomes for the valuable correspondence on the three-link model.  
We thank Georges Harik for many useful discussions. 
We thank Leela Hebbar for help with the manuscript.

\bibliographystyle{unsrtnat}

\bibliography{zerocot_pre2.bbl}

\appendix

\section{Rotation matrix parameterization}
\label{rotmatparam}

The rotations around the axes $X$, $Y$ and $Z$ by an angle $\alpha$ are defined as
\begin{gather}
  R_X(\alpha)=
  \begin{bmatrix}
    1 & 0 & 0 \\
    0 & c & -s \\
    0 & s & c
  \end{bmatrix}
  ,\quad
  R_Y(\alpha)=
  \begin{bmatrix}
    c & 0 & s \\
    0 & 1 & 0 \\
    -s & 0 & c
  \end{bmatrix}
  , \notag \\
  R_Z(\alpha)=
  \begin{bmatrix}
    c & -s & 0 \\
    s & c & 0 \\
    0 & 0 & 1
  \end{bmatrix}
  ,
\end{gather}
where we have defined $c = \cos{\alpha}$ and $s = \sin{\alpha}$. Let $R_{A_i A_{i+1} ... A_j}(\alpha_i, \alpha_{i+1}, ... , \alpha_j)$ be a sequence of intrinsic (i.e. relative to the body-fixed frame) rotations, where $(k+1)$-th rotation is taken around the body axis $A_{i+k} \in \{X,Y,Z\}$ by the angle $\alpha_{i+k}$:
\begin{multline}
  R_{A_i A_{i+1} ... A_j}(\alpha_i, \alpha_{i+1}, ... , \alpha_j) = 
  R_{A_i}(\alpha_i)R_{A_{i+1}}(\alpha_{i+1}) \\ ... R_{A_j}(\alpha_j)
\end{multline}
We will also use shorthand notations $R_i = R_{A_i}(\alpha_i)$ and $R_{i:j}$, defined as
\begin{equation}
  R_{i:j} =
  \left\{
  \begin{split}
    i \le j: & \quad R_i R_{i+1} ... R_j \\
    i = j+1: & \quad I \\
    i > j+1: & \quad {\rm undefined}
  \end{split}
  \right.
\end{equation}
Assume now that $q_{\alpha} \equiv [\alpha_i, \alpha_{i+1}, ... , \alpha_j]^\T$ is a function of time. Let $\hat \omega_i = \dot R_i R_i^\T$ and $\hat \omega_{i:j} = \dot R_{i:j} R_{i:j}^\T$. It is straightforward to verify, for $i < j$: 
\begin{equation}
  \hat\omega_{i:j} 
  = \sum_{k=i}^{k=j} R_{i:k-1} \hat\omega_k R_{i:k-1}^\T
  = \sum_{k=i}^{k=j} \widehat{R_{i:k-1} \omega_k}
  \label{omij}
\end{equation}
 which implies $\omega_{i:j} = \sum_{k=i}^{k=j} R_{i:k-1} \omega_k$. One can check that $\omega_i = I_{(\cdot f(A_i))}\dot\alpha_i$, where $f(A)$ maps $X \to 1$, $Y \to 2$ and $Z \to 3$. Let $R_{i:j} = R_{ab}$ be a rotation that takes $F_a$ to $F_b$. Then $\omega_{i:j}$ is the angular velocity of $F_b$ relative to $F_a$. Let us define $S_{ab}$ to be a transformation relating the generalized coordinate velocity $\dot q_{\alpha}$ to the angular velocity $\omega_{i:j}$, that is $\omega_{i:j} = S_{ab} \dot q_{\alpha}$. We then find:
\begin{multline}
  S_{ab} = \left[I_{(\cdot f(A_i))},\, R_{i(\cdot f(A_{i+1}))},\, R_{i:i+1(\cdot f(A_{i+2}))},\, \right. \\
    \left. ... R_{i:k-1(\cdot f(A_k))}\right]
  \label{genveltransf}
\end{multline}
Note, the order of columns in $S_{ab}$ follows the order of components in $q_{\alpha}$, (as indicated by $A$'s and $\alpha$'s subscripts). In the context of rotation sequences, we will refer to $R_{i:k}$ for $k<j$ as partial rotations. Thus, the matrix $S_{ab}$ is specified in terms of the partial rotations.

The angular velocity $\omega_{i:j}$ in Eq.(\ref{omij}) is specified in $F_a$. To transform $\dot q_\alpha$ to another $F_c$, we include a rotation $R_{ca}$: $\omega_{i:j}^c = R_{ca}\omega_{i:j} = S_{ab}^c \dot q_\alpha$, where we have defined $S_{ab}^c \equiv R_{ca}S_{ab}$. To avoid a confusion, we never omit the superscript for the angular velocity transformation $S_{ab}^c$, unless $c = a$.  
We use the following shortcut notations in the paper: $S_1 \equiv S_{gl}$, $S_t \equiv S_{lt}$, $S_1^a \equiv S_{gl}^a$ and $S_t^a \equiv S_{lt}^a$.

In our model the (fictitious) contact joint parameterization is chosen so that $R_1 = R_{ZYX}(\psi_l,\theta_l,\phi_l)$. The hip joint geometry is chosen so that $R_t = R_{YX}(\theta_t,\phi_t)$. Using Eq.(\ref{genveltransf}), and paying attention to the order of joint angles in $q_l$ and $q_t$, we can write down the expressions for the generalized velocity transformations $S_1$ and $S_t$ corresponding to $R_1$ and $R_t$, as given in Eqs.(\ref{s1},\ref{st}).

\section{Derivation of exact equations of motion.}
\label{exacteqmotderivation}

In this section we derive the representation of the mass matrix $H$ and the potential term $V$ in terms of the model constants and the joint rotations, as given in Sec. \ref{exacteqmot}.

We use shortcut notations $R_1 \equiv R_{gl}$, $R_2 \equiv R_{gt}$, $R_t \equiv R_{lt}$, $\Omega_1 \equiv \Omega_{gl}$, $\Omega_2 \equiv \Omega_{gt}$ and $\Omega_t \equiv \Omega_{lt}$. We call $R_1$, $R_t$, $\Omega_1$ and $\Omega_t$ joint rotations and angular velocities. Clearly, $\Omega_2 = \Omega_1 + \Omega_t$. Since $v_i = \dot r_i$, $\dot r_1 = \hat\Omega_1 r_1$, $\dot r_h = \hat\Omega_1 r_h$, $\dot r_t = \hat\Omega_2 r_t$ and $r_2 = r_h + r_t$, we have:
\begin{equation}
  \begin{split}
    & v_1 = \hat\Omega_1 r_1 \\
    & v_2 = \hat\Omega_1 r_2 + \hat\Omega_t r_t
  \end{split}
  \label{velsviaangvels}
\end{equation}
We can now write the kinetic energy (see Eq.(\ref{kinpotenergy})) in the basis of joint angular velocities $\Omega' = [\Omega_1;\Omega_t]$ as $T = \frac{1}{2}\Omega'^\T H' \Omega'$, where
\begin{equation}
  H' = 
  \begin{bmatrix}
    m_1 \hat r_1^\T \hat r_1 + m_2 \hat r_2^\T \hat r_2 + I_1 + I_2 \, & m_2 \hat r_2^\T \hat r_t + I_2 \\
    m_2 \hat r_t^\T \hat r_2 + I_2 & m_2 \hat r_t^\T \hat r_t + I_2
  \end{bmatrix}
\end{equation}
Let us express $\Omega'$ in terms of $\dot q$. According to App. \ref{rotmatparam}, $\Omega_1^g = S_{gl}^g \dot q_l = S_1 \dot q_l$ and $\Omega_t^g = S_{lt}^g \dot q_t = R_1 S_t \dot q_t$. Therefore $\Omega' = S' \dot q$, where
\begin{equation}
  S' =
  \begin{bmatrix}
    S_1 & 0 \\
    0 & R_1 S_t
  \end{bmatrix}
  .
\end{equation}
Let us rotate $H'$ partwise so that corresponding part's moments of inertia are diagonal in the new basis. In other words, we write $H' = R' \bar H R'^\T$, where
\begin{equation}
  R' =
  \begin{bmatrix}
    R_1 & 0 \\
    0 & R_2
  \end{bmatrix}
\end{equation}
To compute $\bar H = R'^\T H' R'$ we use $r_1^g = R_1 r_1^l$, $r_2^g = R_1 r_h^l + R_2 r_t^t$, $I_1^g = R_1 I_1^l R_1^\T$, $I_2^g = R_2 I_2^t R_2^\T$ and $R_2 = R_1 R_t$ to find $\bar H$ given in Eq.(\ref{barh}). Note that $\bar H$ depends only on the model constants and $R_t$. We can now write $T = \frac{1}{2}\Omega'^\T H' \Omega' = \frac{1}{2}\dot q^\T \bar S^\T \bar H \bar S \dot q = \frac{1}{2}\dot q^\T H \dot q$  where $\bar S = R'^\T S'$ is given in Eq.(\ref{bars}). The expression for $V$ in Eq.(\ref{potenviarelrots}) should be obvious now.

\section{Linear dynamics approximation}
\label{lindynapprox}

In the linear dynamics limit one retains all the terms up to order $\mathcal{O}(q)$ in the equations of motion Eq.(\ref{eqmotion}). This implies that the mass matrix $H$ and the Coriolis term $C$ must be considered up to order $\mathcal{O}(1)$, and the potential term $G$ up to order $\mathcal{O}(q)$. The Coriolis term vanishes at this level, as clear from its definition in Eq.(\ref{coriolis}). We consider $H$ and $G$ below. Note, the derived expressions are exact to all orders in $d$. In this section below we refer to the quantities $H$, $R_1$, $R_t$, $S_1$, $S_t$, $\bar S$, $V$ and $G$ that are defined or used in Eqs.(\ref{massmatrix}-\ref{barh}) and Eqs.(\ref{potenviarelrots},\ref{potterm}). 

\subsection{Mass matrix $H$ in linear dynamics approximation}
\label{massmatlindyn}

We need to compute $H$ to order $\mathcal{O}(1)$ in $q$. To that order, $R_1 = R_t = S_1 = I$, $S_t = I_{3\times 2}$ and $\bar S = [I_{3\times 3},0;0,I_{3\times 2}]$. Using Eq.(\ref{modelrs}) and Eq.(\ref{skewsymprod}) to compute products of the form $\hat r^\T \hat r'$ in Eq.(\ref{barh}), one readily arrives at the expressions below. 

In the linear dynamics limit $H \to H_0 = [H_{ll}^0, H_{lt}^0; H_{tl}^0, H_{tt}^0]$, where
\begin{align}
  & H_{ll}^0 =
  \begin{bmatrix}
    \mu_2 + I_{\phi} + \mu_0 d^2 & 0 & 0 \\
    0 & \mu_2 + I_{\theta} & -\mu_1 d \\
    0 & -\mu_1 d & I_{\psi} + \mu_0 d^2
  \end{bmatrix}
  , \notag \\
  & H_{tl}^0 =
  \begin{bmatrix}
    -m_2 l_2 l_t + I_{2\phi} & 0 & 0 \\
    0 & -m_2 l_2 l_t + I_{2\theta} & m_2 l_t d
  \end{bmatrix}
  , \notag \\
  & H_{tt}^0 =
  \begin{bmatrix}
    m_2 l_t^2 + I_{2\phi} & 0 \\
    0 & m_2 l_t^2 + I_{2\theta}
  \end{bmatrix}
  ,
\end{align}
and $H_{lt}^0 = {H_{tl}^0}^\T$.

\subsection{Potential term $G$ in linear dynamics approximation}

To compute $G$ to order $\mathcal{O}(q)$ we need to compute the potential energy $V$, and hence the joint rotation matrices $R_1$ and $R_t$, (see Eq.(\ref{potenviarelrots})), to order $\mathcal{O}(q^2)$. To quadratic order we find:
\begin{multline}
  R_1 = R_{ZYX}(\psi,\theta,\phi) \\ = 
  \begin{bmatrix}
    1-\frac{\theta^2 + \psi^2}{2} & -\psi & \theta \\
    \psi & 1-\frac{\phi^2 + \psi^2}{2} & -\phi \\
    -\theta & \phi & 1-\frac{\phi^2 + \theta^2}{2}
  \end{bmatrix}
\end{multline}
and
\begin{equation}
  R_t = R_{YX}(\theta,\phi) = 
  \begin{bmatrix}
    1-\frac{\theta^2}{2} & \phi\theta & \theta \\
    0 & 1-\frac{\phi^2}{2} & -\phi \\
    -\theta & \phi & 1-\frac{\phi^2 + \theta^2}{2}
  \end{bmatrix}
  .
\end{equation}
From Eq.(\ref{potenviarelrots}) we find to quadratic order:
\begin{multline}
  V = \mu_1 + \mu_0 d\phi_l -\mu_1\frac{\phi_l^2 + \theta_l^2}{2} \\
  + m_2 l_t \left(\frac{\phi_l^2 + \theta_l^2}{2} 
  + \phi_l\phi_t + \theta_l\theta_t \right).
\end{multline}
For $G$ we find $G = G_0 + G_1 q$, where $G_0 = [g\mu_0 d, 0, 0, 0, 0]^\T$ and
\begin{equation}
  G_1 =
  -g\mu_1 
  \begin{bmatrix}
    1 & 0 & 0 & 0 & 0 \\
    0 & 1 & 0 & 0 & 0 \\
    0 & 0 & 0 & 0 & 0 \\
    0 & 0 & 0 & 0 & 0 \\ 
    0 & 0 & 0 & 0 & 0 
  \end{bmatrix}
  + gm_2l_t 
  \begin{bmatrix}
    0 & 0 & 0 & 1 & 0 \\
    0 & 0 & 0 & 0 & 1 \\
    0 & 0 & 0 & 0 & 0 \\
    1 & 0 & 0 & 1 & 0 \\
    0 & 1 & 0 & 0 & 1 
  \end{bmatrix}
  .
\end{equation}

\section{Higher order correction to linear dynamics approximation}
\label{nextorder}

The next order terms on top of the linear dynamics approximation captured by Eq.(\ref{hg0g1alpha}), come from expanding $H_0$ to order $\mathcal{O}(\bar{d})$, and from expanding $H$ and $C$ to order $\mathcal{O}(q)$ at $d=0$. The latter constitutes a nonlinear correction to the linear dynamics approximation. We denote those terms $H_1(q)$ and $C_1(\dot q)$ for the mass matrix and the Coriolis term corrections respectively. Expanding $H_0$ (defined in App. \ref{massmatlindyn}) to order $\mathcal{O}(\bar{d})$ contributes the term $db^\T\ddot q_\theta$ in Eq.(\ref{eqmotpsi}). For $H_1(q)$ one finds:
\begin{equation}
  H_1(q) = e_3 q^\T A + A^\T q e_3^\T,
\end{equation}
where $e_3 = [0,0,1,0,0]^\T$ and $A$ is defined in terms of the elements $A_{\alpha\beta}^{ij} \equiv A_{\alpha\beta(ij)}$ of the matrices $A_{\phi\theta}$ and $A_{\theta\phi}$, given in Eq.(\ref{aptatp}):
\begin{equation}
  A =
  \begin{bmatrix}
    0 & A_{\phi\theta}^{11} & 0 & 0 & A_{\phi\theta}^{12} \\
    A_{\theta\phi}^{11} & 0 & 0 & A_{\theta\phi}^{12} & 0 \\
    0 & 0 & 0 & 0 & 0 \\
    0 & A_{\phi\theta}^{21} & 0 & 0 & A_{\phi\theta}^{22} \\
    A_{\theta\phi}^{21} & 0 & 0 & A_{\theta\phi}^{22} & 0
  \end{bmatrix}
\end{equation}
The term $C_1(\dot q)$ is determined by $H_1(q)$ via Eqs.(\ref{coriolis},\ref{christoffel}). We find:
\begin{equation}
  \frac{\partial H_1(q)}{\partial q_{(k)}} = 
  e_3 A_{(k \cdot)} + \left( A^\T \right)_{(\cdot k)} e_3^\T,
\end{equation}
the Christoffel symbols:
\begin{multline}
  \Gamma^k = \frac{1}{2}\bigl( 
  e_3 \left(A + A^\T \right)_{(k \cdot)} -
  \left(A - A^\T \right)_{(\cdot k)} e_3^\T  
  \\ 
  - \left(A - A^\T \right) e_{3(k)}
  \bigr),
\end{multline}
and the Coriolis term:
\begin{multline}
  C_1(\dot q) = \frac{1}{2}\left(
  e_3 \dot q^\T \left(A + A^\T \right) 
  \right. \\ \left. - 
  \left(A - A^\T \right)\left( \dot q e_3^\T + e_3^\T \dot q I_{5 \times 5} \right)
  \right)
  .
\end{multline}
The corresponding contribution to the equations of motion Eq.(\ref{eqmotpsi}) in the $\psi$ sector is
\begin{multline}
  e_3^\T \left(H_1(q) \ddot q + C_1(\dot q) \dot q \right) 
  = q^\T A \ddot q + \frac{1}{2}\dot q^\T \left( A + A^\T \right) \dot q 
  \\
  = \frac{d}{dt}\left( q^\T A \dot q \right) 
  = \frac{d}{dt}\left( q_\phi^\T A_{\phi\theta} \dot q_\theta 
  + q_\theta^\T A_{\theta\phi} \dot q_\phi \right).
\end{multline}

\section{Coronal sector impact time equations}
\label{coronalimptimes}

To reduce clutter, we omit the coronal sector superscript $\phi$ in this section. The last four equations in Eq.(\ref{coronaljoinconds}) form a homogeneous linear system $Aw = 0$, where
\begin{equation}
  A = 
  \begin{bmatrix}
    u_{+(1)}^s \omega_+^s s_+^s \quad & u_{-(1)}^s \omega_-^s s_-^s & 0 \\
    u_{+(1)}^s {\omega_+^s}^2 c_+^s \quad & -u_{-(1)}^s {\omega_-^s}^2 c_-^s & 0 \\
    u_{+(2)}^s c_+^s \quad & u_{-(2)}^s c_-^s & s^d \\
    u_{+(2)}^s \omega_+^s s_+^s \quad & u_{-(2)}^s \omega_-^s s_-^s & -\omega^d c^d
  \end{bmatrix}
  ,
\end{equation}
where we have defined $c_z^s = \cosh{\sqrt{z}\omega_z^s t_s}$, $s_z^s = \sqrt{z}\sinh{\sqrt{z}\omega_z^s t_s}$, $c^d = \cos{\omega^d t_d}$ and $s^d = \sin{\omega^d t_d}$.

Consider the task of reducing the rank of a $n\times m$ matrix $M$, with $n > m$, below $m$. To achieve it, one needs to tune at least $n-m+1$ parameters of $M$, in general. Indeed, for $A$ this can be done by tuning two parameters, $t_s$ and $t_d$. 

Consider two submatrices of $A$: 1) a $2\times 2$ matrix $A_1$, which is the top left corner submatrix of $A$, 2) a $3\times 3$ matrix $A_2$, which is obtained from $A$ by dropping the second row. If $A_1$ and $A_2$ are singular matrices, then $A$ has rank at most 2. One can see it as follows. The singularity of $A_1$ implies that the rows $A_{(1\cdot)}$ and $A_{(2\cdot)}$ are not independent. The singularity of $A_2$ implies that among $A_{(1\cdot)}$, $A_{(3\cdot)}$ and $A_{(4\cdot)}$ there are at most two independent rows. Hence $A$ has at most two independent rows.

Let $A{:}A_{(i\cdot)}$ be a matrix obtained from $A$ by the element-wise division of its rows by $i$-th row. We first test the singularity of $A_1$:
\begin{equation}
  \det{(A_1{:}A_{1(1\cdot)})} = 
  \omega_-^s \cot{\omega_-^s t_s} - \omega_+^s \coth{\omega_+^s t_s},
  \label{deta1}
\end{equation}
from where the first equation in Eq.(\ref{tstdcoronal}) follows.

To test the singularity of $A_2$ we compute:
\begin{align}
  & \det{(A_2{:}A_{2(3\cdot)})} = 
  \label{deta2} \\ 
  & = 
  -\frac{\sigma_+}{\omega_-^s} \cot{\omega_-^s t_s}
  -\frac{\sigma_-}{\omega_+^s} \coth{\omega_+^s t_s}
  +\frac{\sigma_+-\sigma_-}{\omega^d} \tan{\omega^d t_d}
  \notag \\ 
  & = 
  -\left(\frac{\sigma_+}{{\omega_-^s}^2}+\frac{\sigma_-}{{\omega_+^s}^2}\right)
  \omega_-^s \cot{\omega_-^s t_s} 
  +\frac{\sigma_+-\sigma_-}{\omega^d} \tan{\omega^d t_d},
  \notag
\end{align}
where $\sigma_z = u_{z(1)}^s / u_{z(2)}^s$. In the second step we assumed $\det{A_1}=0$ and used Eq.(\ref{deta1}). Next, we compute
\begin{multline}
  \frac{\sigma_+-\sigma_-}{\frac{\sigma_+}{{\omega_-^s}^2}+\frac{\sigma_-}{{\omega_+^s}^2}} 
  = \frac{\Lambda_{-,+}^\lambda-\Lambda_{-,-}^\lambda}{\Lambda_{-,-}^\lambda\Lambda_{+,+}^\lambda - \Lambda_{-,+}^\lambda\Lambda_{+,-}^\lambda} \\
  = \frac{1}{\Lambda_{-,+}^\lambda - \Lambda_{+,+}^\lambda}
  = \frac{1}{-a_-} = {\omega^d}^2
  \label{soratio}
\end{multline}
where we used Eq.(\ref{eigensystem}) and Eq.(\ref{lambdauseful}). From Eq.(\ref{deta2}) and Eq.(\ref{soratio}) the second equation in Eq.(\ref{tstdcoronal}) follows.

\section{Model parameterization}
\label{modparamapp}

In this section we fill in the details omitted in Sec. \ref{modparam}.
Our goal is to realize the impact phase solution (at a given $\bar\tau$) by tuning the minimal number of the model parameters. We first focus on the $\phi$ sector. The impact phases are completely specified by $a_z^\phi$, $\gamma$ and $\tau$. Since the impact phases are defined by three frequencies, we are going to tune $\tau$ and two model parameters. To this end, we first solve for $a_+^\phi$ and $\gamma$, with $a_-^\phi$ fixed and $\tau = \bar\tau\sqrt{\gamma}$. Let us write Eq.(\ref{azfromlambdaz}) as $a_z^\phi = \gamma \bar{a}_{+z}^\phi = \gamma \Lambda_{+z}(\bar\lambda_+^\phi, \bar\lambda_-^\phi, 1)$. From where we get: $a_+^\phi = a_-^\phi \bar{a}_{++}^\phi / \bar{a}_{+-}^\phi$ and $\gamma^2 = {a_-^\phi}^2 ({\bar{a}}_{+-}^\phi)^{-2}$. Using Eq.(\ref{lambdauseful1}) we find
\begin{equation}
  \frac{\bar{a}_{++}^\phi}{\bar{a}_{+-}^\phi} 
  = \frac{\bar{a}_{++}^\phi \bar{a}_{+-}^\phi}{\bar{a}_{+-}^\phi \bar{a}_{+-}^\phi}
  = \frac{\bar\lambda_+^\phi \bar\lambda_-^\phi - 1}{\left(\bar{a}_{+-}^\phi\right)^2}
\end{equation}
We can now write the equations on $a_+^\phi$ and $\gamma^2$ in the form:
\begin{equation}
  \begin{split}
    & a_+^\phi = -a_-^\phi \left(\kappa_1^2+1\right)\kappa_2 \\
    & \gamma^2 = {a_-^\phi}^2 \kappa_2
  \end{split}
  \label{azgammakappas}
\end{equation}
where $\kappa_i$ were defined in Eq.(\ref{kappa12}). Let us work in dimensionless units by setting $g=m_2=l_t=1$. Let us also parameterize the moments of inertia as $I_{1\alpha} = c_1^\alpha \tilde\mu_2$ and $I_{2\alpha} = c_2^\alpha m_2l_t^2 = c_2^\alpha$. The expressions for $a_z^\phi$ and $\gamma^2$, in terms of the model parameters (see Eq.(\ref{azbetagamma})), now read: 
\begin{equation}
  a_+^\phi = \frac{\tilde\mu_2}{\tilde\mu_1} \left(1+c_1^\phi\right), \,\,\,
  a_-^\phi = -\left(1+c_2^\phi\right), \,\,\,
  \gamma^2 = \frac{l_h^2}{\tilde\mu_1}.
  \label{azgammamodel}
\end{equation}
We first consider a simpler case of $c_i^\phi = 0$. Plugging $a_z^\phi$ and $\gamma^2$ from Eq.(\ref{azgammamodel}) into Eq.(\ref{azgammakappas}) and solving for $l_h$ and $l_1$, one readily finds Eq.(\ref{lhl1}). We next consider the general case of $c_i^\phi > 0$. Let us express $\kappa_i$ as
\begin{equation}
  \begin{split}
    & \kappa_1^2+1 = \left({\kappa'_1}^2+1\right)
    \left(1+c_1^\phi\right)\left(1+c_2^\phi\right) \\
    & \kappa_2 = \kappa'_2\left(1+c_2^\phi\right)^{-2}
  \end{split}
  \label{kappac}
\end{equation}
If we now plug $a_z^\phi$ and $\gamma^2$ from Eq.(\ref{azgammamodel}) and $\kappa_i$ from Eq.(\ref{kappac}) into Eq.(\ref{azgammakappas}), we get equations on $l_h$ and $l_1$ which look exactly like $c_i^\phi = 0$ case, but with $\kappa_i$ replaced by $\kappa'_i$. It follows that the general case solution must have the form of Eq.(\ref{lhl1}) with $\kappa_i$ replaced by $\kappa'_i$.

We consider the $\theta$ sector next. Let us express $a_-^\theta$ in terms of $a_+^\theta$, $\lambda_-^\theta$ and $\gamma^2$, by inverting $\lambda_-^\theta = \lambda_{+-}^\theta$, see Eq.(\ref{eigensystem}). We find:
\begin{equation}
  a_-^\theta = \lambda_-^\theta + \frac{\gamma^2}{\lambda_-^\theta - a_+^\theta}
  \label{amtheta}
\end{equation}
Eq.(\ref{osagittal}) implies $\lambda_-^\theta = -\tau^2/\pi^2$. Notice that $\tilde\mu_2 / \tilde\mu_1$ and $\gamma^2$ are purely functions of $\kappa'_i$: $\tilde\mu_2 / \tilde\mu_1 = ({\kappa'_1}^2+1)\kappa'_2$ and $\gamma^2 = \kappa'_2$. Therefore we can write $a_+^\theta = ({\kappa'_1}^2+1)\kappa'_2 (1+c_1^\theta)$. Plugging the above expressions for $\lambda_-^\theta$, $\gamma^2$ and $a_+^\theta$ into Eq.(\ref{amtheta}), while replacing $a_-^\theta$ with $-(1+I_{2\theta})$, we obtain Eq.(\ref{i2thetagen}) and the restricted version Eq.(\ref{i2thetarestr}).

Let us analyze the function $\Phi(\xi,\kappa_1,\kappa_2)$, implicitly introduced via Eq.(\ref{c1theta}) in Sec. \ref{gencase}. We will use the restricted solution quantities $\tau^r$ and $I_{2\theta}^r$, as indicated by the superscript `r'. Note, $\xi = (\tau^r/\pi)^2$. Using Eq.(\ref{i2thetarestr}), we can cast $\Phi(\xi,\kappa_1,\kappa_2)$ in the form:
\begin{multline}
  \Phi(\xi,\kappa_1,\kappa_2) \equiv 
  \frac{(1-\xi^2)^{-1}-\xi^2\kappa_2^{-1}}{\kappa_1^2+1} \\ = 1 + I_{2\theta}^r
  \frac{\xi^2\kappa_2^{-1}+\kappa_1^2+1}{(1-\xi^2)\left(\kappa_1^2+1\right)}.
\end{multline}
Since $I_{2\theta}^r$ and $\kappa_2$ are both positive, $\Phi(\xi,\kappa_1,\kappa_2) > 1$ provided $\xi < 1$.

To prove that $0 < u_{-(1)}^\theta < u_{-(2)}^\theta$, we observe that $0 < \lambda_{--} \le \gamma < \gamma\beta$, since $\tilde\mu_1 > m_2 l_t$, see Eq.(\ref{azbetagamma}). The result then follows from Eq.(\ref{eigensystem}). Also, $u_{-(2)}^\theta > \beta u_{-(1)}^\theta$.

\section{Up-down torso duality}
\label{dualityapp}

In the dimensionless units, $H_\alpha$ and $G_{1\alpha}$ are
\begin{equation}
  H_\alpha = 
  \begin{bmatrix}
    h_1 & h_z \\
    h_z & h_2
  \end{bmatrix}
  , \quad
  G_{1\alpha} =
  \begin{bmatrix}
    g_1 & z \\
      z & z + \ke
  \end{bmatrix}
  ,
\end{equation}
where
\begin{equation}
  \begin{split}
    & h_1 = \mu_2 + I_\alpha, 
    \quad h_2 = 1 + I_{2\alpha},
    \quad h_z = zl_h + h_2, \\
    & g_1 = -\mu_1.
  \end{split}
\end{equation}
Computing $\tilde H_\alpha$ and $\tilde G_{1\alpha}$ (see Eq.(\ref{tHatG1a})) with $\tilde S$ defined in Eq.(\ref{tS}) and selecting $\y$ that diagonalizes $\tilde G_{1\alpha}$ we find
\begin{equation}
  \begin{split}
    & \tilde H_\alpha = 
    \begin{bmatrix}
      h_1-2zh_z\y+h_2y^2  & zh_z-h_2y \\
      zh_z-h_2y & h_2
    \end{bmatrix}
    , \\
    & \tilde G_{1\alpha} =
    \begin{bmatrix}
      g_1 -\y & 0 \\
      0 & \frac{1}{\y}
    \end{bmatrix}
    ,
  \end{split}
  \label{thtg1z}
\end{equation}
where $\y = 1/(z+\ke)$. As explained in Sec. \ref{duality}, we want $\tilde H_\alpha$ and $\tilde G_{1\alpha}$ to be independent of $z$. It then follows from Eq.(\ref{thtg1z}) that $\y$, $g_1$, $h_1$, $h_2$ and $zh_z$ are all independent of $z$. Solving $\y' = \y$ and $z'h_z' = zh_z$ with $z' = -z$, where the prime denotes quantities in the dual setup, we find:
\begin{equation}
  \ke' = \ke + 2z, \quad l_h' = l_h + 2zh_2.
\end{equation}
Note that $\mu_1' = \mu_1$. If we demand $I_{1\alpha}$ to be fixed, then we also have $\mu_2' = \mu_2$. From there, the transformation law for $m_1$ and $l_1$ in Eq.(\ref{dualtransf}) easily follows, where $\delta_n = m_1'l_1'^n = \mu_n - (l_h'-z')^n$. To derive a condition for $\delta_n > 0$, we use Eq.(\ref{lhl1}) to substitute $l_1$ and $l_h$ to find a quadratic equation on $\sqrt{m_1}$. Solving the quadratic equations we arrive at Eq.(\ref{mc12}). The condition in Eq.(\ref{mc3}) is similarly derived from $\beta > 2+\y$, which implies $\x < 1/(2+\y)$.

\section{Constrained gradient descent}
\label{constrgraddesc}

We enforce the triangle inequality (for the moments of inertia) by setting up a one-to-one correspondence between triangle's side lengths $(d_1,d_2,d_3)$ and a triplet $(\eta_1,\eta_2,\eta_3)$ of real numbers. We require that the triangle's perimeter is $\sum_i d_i = P = \sum_i e^{\eta_i}$ and the opposing angles are $\alpha_i = \pi e^{\eta_i}/P$. One then finds for the mapping:
\begin{equation}
  \begin{split}
    & d_i = \frac{P \sin{\frac{\pi e^{\eta_i}}{P}}}{\sum_i \sin{\frac{\pi e^{\eta_i}}{P}}}, \\
    & \eta_i = \ln{\left(\frac{P}{\pi}
      \arccos{\frac{\sum_{j\ne i}d_j^2 - d_i^2}{2\Pi_{j\ne i} d_j}}\right)}.
  \end{split}
\end{equation}
This parameterization is only suitable for enforcing the strict triangle inequality (the general case in Sec. \ref{gencase}), as $\eta_i \to -\infty$ for $d_i \to 0$.

Consider a constrained optimization problem of finding $x$ that maximizes an objective function $f(x)$ under the constraints $\lVert x-x_0 \rVert = \epsilon$ and $h(x) = 0$. Solving it amounts to finding the stationary points of the Lagrangian function
\begin{equation}
  f(x) + \lambda_h^\T h(x) + \frac{\lambda_x}{2}
  \left((x-x_0)^\T (x-x_0) - \epsilon^\T\epsilon \right)
\end{equation}
with respect to $x$ and the Lagrange multipliers $\lambda_h$ and $\lambda_x$. Assuming that $x_0$ approximately satisfies $h = 0$ and $\epsilon$ is small, we can optimize an approximate Lagrangian function where $f(x)$ and $h(x)$ have been replaced by their first order expansion around $x_0$. The result of this is Eq.(\ref{cgd}) with implicit dependence of $\lambda_x$ on $\epsilon$, (with $x_0$ and $x$ replaced by $x_i$ and $x_{i+1}$). In practice, we dispense with $\epsilon$ and treat $\lambda_x$ as the controlling parameter of the gradient descend rate.

\section{Miscellaneous relations and properties}
\label{miscrels}

In this section we gather an eclectic mix of relations and properties used in the paper.

Product of skew symmetric matrices can be expressed via respective vectors as
\begin{equation}
  \hat a^\T \hat b = (a^\T b) I - b a^\T
  \label{skewsymprod}
\end{equation}

It is easy to see that a real-valued $2\times 2$ matrix $M$ with $\det{M} < 0$ has one positive and one negative eigenvalues: the discriminant of the characteristic polynomial $\lambda^2 - \lambda{\rm Tr}{M} +\det{M}$ is positive, therefore the eigenvalues are real, nonzero and of different sign.

It is easy to verify a number of useful relations on $\Lambda_{zz'}$ using its definition in Eq.(\ref{Lambda}):
\begin{equation}
  \begin{split}
    & \Lambda_{z,z'}(a,b,c) + \Lambda_{-z,-z'}(a,b,c) = a \\
    & \Lambda_{z,z'}(a,b,c) - \Lambda_{-z,z'}(a,b,c) = zb
  \end{split}
  \label{lambdauseful}
\end{equation}
and
\begin{equation}
  \begin{split}
    & \Lambda_{-+}(a,b,c) \Lambda_{--}(a,b,c) = -c \\
    & \Lambda_{++}(a,b,c) \Lambda_{+-}(a,b,c) = ab-c
  \end{split}
  \label{lambdauseful1}
\end{equation}

To prove Eq.(\ref{alambdagamma}) relating the coronal sector frequencies, we compute
\begin{equation}
  (a_- - \lambda_-)(a_- - \lambda_+) = 
  \lambda_{--}\lambda_{-+} = \gamma^2
\end{equation}
where we used Eqs.(\ref{lambdazzazz},\ref{eigensystem},\ref{lambdauseful},\ref{lambdauseful1}).

The values of $\chi$, that is the initial conditions on $q^p$ and $t_p$, fully specifying (in addition to the model parameters $\eta$) the (3,4)-mode, numerically studied in Sec. \ref{gensol}, are given below:
\begin{equation}
  \begin{split}
    & \phi^s_l = 0.2146047,\\
    & \phi^s_t = -0.1452115,\\
    & \dot\theta^s_l = 0.5255085,\\
    & \dot\psi^s_l = 0.2670555,\\
    & \dot\theta^s_t = 0.1306085\\
    & \dot\theta^d_l = 0.5369450,\\
    & \dot\phi^d_t = 0.1716506,\\
    & \dot\theta^d_t = 0.1257789,\\
    & t_s = 1.1657382,\\
    & t_d = 1.1582562.
  \end{split}
\end{equation}

It is straightforward to compute the minimum friction coefficient $\rho$ given the moment of gravity force $M_g = \sum_i m_i \hat r_i g_f$, the momentum $P = \sum_i m_i v_i$ and the angular momentum $L = \partial\mathcal{L}/\partial\Omega_1$ \cite{landau2000mechanics}. One readily finds $M_g = R_1 (m_1 \hat r_1^l + m_2 \hat r_2^l)g_f$ and $P = R_1 [m_1 \hat r_1^l + m_2 \hat r_2^l; m_2 \hat r_t^t R_t^\T]^\T\bar{S}\dot q$, where $r_2^l = r_h^l + R_t r_t^t$. It is also easy to find $L = R_1(\bar{H}\bar{S}\dot q)_{(l)}$, see the last paragraph of App. \ref{exacteqmotderivation}.

\end{document}